\documentclass[%
 reprint,
 prb,
 amsmath,amssymb,
 aps
]{revtex4-2}
\usepackage{hyperref}
\usepackage{hyperref}
\usepackage{graphicx}
\usepackage{bm}
\usepackage{enumitem}
\usepackage{amsmath}
\usepackage{amsfonts}
\usepackage{amssymb}
\usepackage{amsthm}
\begin{document}
\preprint{APS/123-QED}
\title{Spin-polarization of the electric current in half-metallic Co$_2$MnSi Heusler thin films}
\author{J. Solano$^{1,2}$}
\email{jose.solano@fzu.cz}
\author{A. Friedel$^{3,4}$}
\author{Q. Rossi$^{1}$}
\author{J. Robert$^{1}$}
\author{Y. Henry$^{1}$}
\author{P. Pirro$^{4}$}
\author{S. Petit-Watelot$^{3}$}
\author{S. Andrieu$^{3}$}
\author{M. Bailleul$^{1}$}
\email{matthieu.bailleul@ipcms.unistra.fr}
\affiliation{$^{1}$ Institut de Physique et Chimie des Matériaux de Strasbourg, UMR 7504 CNRS, Université de Strasbourg, 23 rue du Loess, BP 43, 67034 Strasbourg Cedex 2, France}
\affiliation{$^{2}$ Institute of Physics, Czech Academy of Sciences, Cukrovarnická 10, 16200 Praha 6, Czech Republic}
\affiliation{$^{3}$ Université de Lorraine, CNRS, IJL, F-54000 Nancy, France}
\affiliation{$^{4}$ Fachbereich Physik and Landesforschungszentrum OPTIMAS, Rheinland-Pfälzische Technische Universität Kaiserslautern-Landau, 67663 Kaiserslautern, Germany}

\begin{abstract}
Using propagating spin wave spectroscopy we measure the spin wave Doppler shift in patterned MgO/Co$_2$MnSi/MgO thin films and determine the degree of spin-polarization of the electric current. Our measurements reveal that the current is fully spin-polarized in the devices. This shows that the half-metallic character of the electron band structure translates into a fully spin polarized current flowing across the patterned films. Additionally, we measure a current-induced change of the spin-wave attenuation from which we estimate the non-adiabatic spin-transfer-torque parameter.

\end{abstract}
\maketitle
\section{\label{sec:level1_intro}Introduction\protect}
A spin-polarized electron band structure is the defining property of any ferromagnet, yet only half-metals are predicted to have an actual 100$\%$ spin-polarization due to the presence of a gap at the Fermi level of one of its spin-bands \cite{Groot.1983}. A text book example of such materials is the Heusler compound Co$_2$MnSi, which has a 0.4-0.8 eV spin-gap at the Fermi level \cite{Andrieu.2016, Guillemard.2019}. The fact that there are only majority electron states in such a wide gap, reduces electron-electron interactions resulting in a record low magnetic damping between metallic ferromagnets down to $\alpha$=4.6-9$\times10^{-4}$ \cite{Guillemard.2019,Melo.2021}.

These properties make Co$_2$MnSi promising for applications in spintronics and magnonics \cite{Guillemard.2019, Mantion.2024}. Indeed, the predicted spin gap has been confirmed to exist even at room temperature \cite{Guillemard.2019}. This suggest that only majority electrons should transport the electric current with a 100 $\%$ spin-polarization leading to a relatively simple description of electron transport \cite{Melo.2021}. However, so far no direct measurement of the spin-polarization of the electron current has been made in these films, and the bulk half-metallicity can only be assumed from the 100$\%$ spin-polarized density of states close to the film surface, as deduced from spin-polarized photoemission experiments\cite{Andrieu.2016, Guillemard.2019}. Moreover, the translation from band structure to electron transport is not direct, as electron scattering with phonons, magnons, defects, and the presence of spin-orbit interaction could lead to a change in spin-polarization of the electron current \cite{Solano.2025}. Therefore, experimentally determining that the electric current remains fully spin-polarized at room temperature in patterned Co$_2$MnSi films would confirm its potential as a key material for spintronic applications. 

In this work, we inject an electric current in MgO/Co$_2$MnSi/MgO thin films where spin waves propagate and measure the spin wave Doppler frequency shifts from which we determine the degree of spin-polarization of the electron current. We show that in agreement with the half-metallicity of the films, the diffusive electric current is indeed 100$\%$ spin-polarized at room temperature, even after patterning. Our technique also determines the current-induced attenuation of the spin waves from which we estimate the non-adiabatic component of the spin-transfer-torque process.

\section{\label{sec:level2_Model}
Films and devices}
\begin{figure}[ht]
\includegraphics[width=0.48\textwidth]{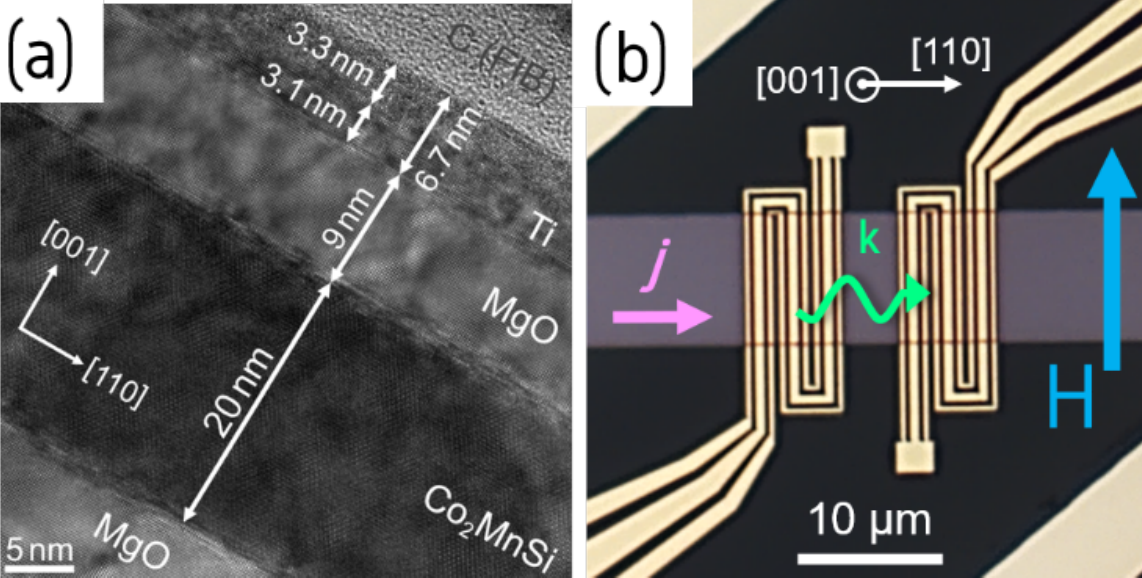}%
\caption{\label{fig:Device} (a) High-resolution transmission electron microscopy micrograph of the deposited film stack. (b) Microscope picture of a spin wave Doppler shift device. $\mathbf{j}$ is the applied DC current, $\mathbf{k}$ is the wave vector, $\mathbf{H}$ the applied magnetic field.}
\end{figure}

The films are grown on commercial MgO substrates by molecular beam epitaxy following a process similar to that detailed elsewhere \cite{Friedel.2025, Andrieu.2016, Guillemard.2019} (see also supplemental material \cite{Supplemental}). The samples for this study consist of the following stack MgO(001)/MgO(10nm)/Co$_2$MnSi(20nm)/MgO(20nm)/ MgO(8nm)/Ti(4nm). The films were characterized in-situ by reflection high-energy electron diffraction and ex-situ by X-ray diffraction and transmission electron microscopy (TEM) showing good epitaxy and crystallinity comparable to the ones reported before \cite{Guillemard.2019}. In Fig.~\ref{fig:Device}~(a) we present a high-resolution TEM micrograph of the stack obtained from a lamella cut by focused ion-beam. We also confirmed that the chemical order predominantly corresponds to the perfectly ordered structure L2$_1$ \cite{Gaier.2008,Guillemard.2020, Supplemental}. Additionally, we measured the electric and magnetic properties of the films (at 300 K) before and after the subsequent nanofabrication without noticing any significant change: same resistivity ($\rho=42.5\:\mu\Omega cm$) \footnote{In agreement with resistivity \cite{Melo.2021} if one considers their Au capping}, ferromagnetic resonance frequencies, magnetic damping ($\alpha=1.2\times10^{-3}$) and saturation magnetization ($\mu_0$M$_s$=1.33 T) (see details in supplemental material \cite{Supplemental} and refs. \cite{Amorim.2021, Solano.2022} therein).

We patterned the films into strips (along Co$_2$MnSi's [110] direction) for the current conduction and for the propagation of spin waves. DC contacts are patterned to inject the electric current \cite{Supplemental}. The devices are covered with an 85 nm insulating layer of sputtered SiO$_2$. Finally, a pair of antennas [Ti(5nm)/Au(140nm)] are patterned on top of the strip. These are used to excite and measure inductively coherent propagating spin waves \cite{Vlaminck.2010} [Fig.~\ref{fig:Device} (b)].

\section{\label{sec:Experiment}
Experiment and results}
\begin{figure}[ht]
\includegraphics[width=0.48\textwidth]{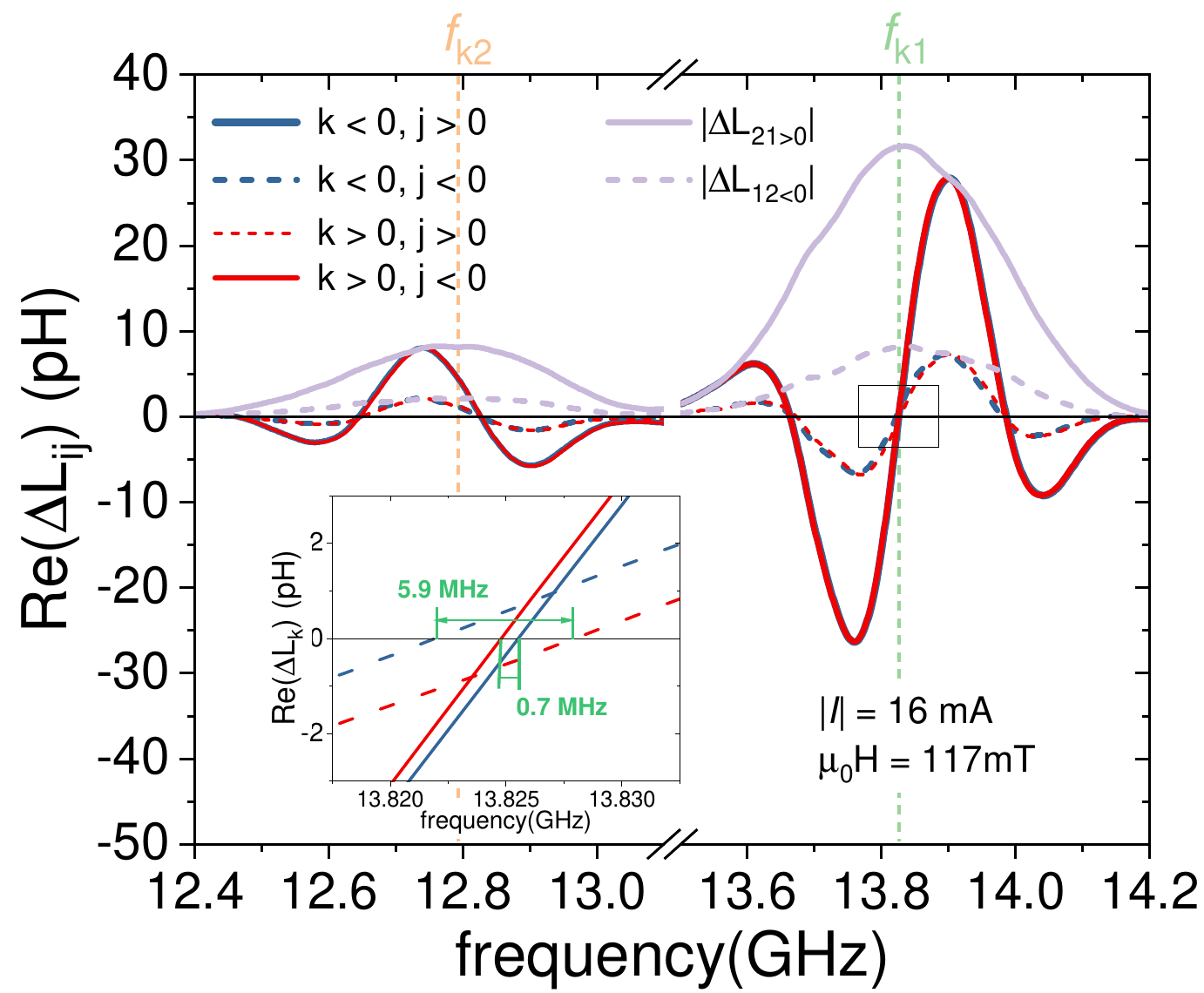}%
\caption{\label{fig:dL} Mutual inductance as function of frequency for spin waves with $k_1$= 3.90 rad/$\mu$m and $k_2$= 1.86 rad/$\mu$m at an applied current $|I|=16 mA$ and magnetic field $\mu_0H=117 \text{mT}$. The inset shows a closer look of the current-induced frequency shifts at $f_{k_1}$.}
\end{figure}

The antennas are connected via two RF probes to a vector network analyzer that provides a microwave current to measure the two-port scattering matrix between them. The propagating spin wave spectroscopy experiments are all performed at room temperature. From this we extract the mutual inductance $L_{ij}$ ($i,j$=1,2) between the two antennas. For our measurements, the external magnetic field $H$ is applied in-plane perpendicularly to the strip, parallel or anti-parallel to Co$_2$MnSi's [1-10] direction, (Damon-Eshbach geometry, Fig.~\ref{fig:Device}) while we sweep the microwave frequency. When the resonance condition is matched, the excitation magnetic field produced by one antenna can efficiently excite coherent spin waves that propagate until the region bellow the second antenna where their stray field is picked up inductively. This is measured as a change in the mutual inductances $\Delta L_{ij}$ (Fig.~\ref{fig:dL}). The geometry of the antennas is optimized for the excitation of spin waves with wave vectors $k_1$= 3.90 rad/$\mu$m and $k_2$= 1.86 rad/$\mu$m leading to the observation of two main resonance frequencies $f_{k_1}$ and $f_{k_2}$, respectively (see details in \cite{Supplemental}). 

Now, when we inject a DC electric current $\mathbf{j}$ into the Co$_2$MnSi strip, the carriers also transport a spin current due to the ferromagnetic nature of the material. The relative motion between this itinerant spin (or magnetic moment) with respect to the propagating spin waves produces the so called spin-transfer-torque (STT). This torque modifies the linearized equation of motion of the magnetization \cite{Bailleul.2017}:

\begin{equation}
\begin{aligned}
    \label{eq:LLGSTT}
    i \omega \textbf{m} = & \gamma \mu_o(  \mathbf{H_\text{eq}} \times \mathbf{m} -  \mathbf{M_\text{s}} \times \textbf{h}) + i\frac{\alpha}{M_\text{s}}\omega  \mathbf{M_\text{s}} \times\textbf{m} \\ 
    & + i\delta\omega_{Dop}\mathbf{m} - i\frac{\beta}{M_\text{s}} \delta\omega_{Dop}\mathbf{M_\text{s}} \times \textbf{m}.
\end{aligned}
\end{equation}

Here, $\omega$ is the angular frequency, $\gamma$ is the gyromagnetic ratio and $\mu_0$ is the permeability of vacuum. $\textbf{M}_{\text{eq}}$ ($\textbf{H}_{\text{eq}}$) and $\textbf{m}$ ($\textbf{h}$) are the static and dynamic components of the magnetization (effective magnetic field), respectively, $\alpha$ is the Gilbert damping parameter representing magnetic losses. The last two terms in Eq.~\eqref{eq:LLGSTT} correspond to the adiabatic and non-adiabatic parts of the STT, respectively. The first term proportional to $\mathbf{m}$ represents the current-induced shift of the precession frequency; also known as the spin-wave Doppler shift \cite{Vlaminck.2008} 
\begin{equation}
\begin{aligned}
    \label{eq:DopplerShift}
    \delta \omega_\text{Dop} = - \frac{\mu_\text{B}}{ e M_\text{s}} P \: \mathbf{j} \cdot \mathbf{k} = \mathbf{u} \cdot \mathbf{k}.
\end{aligned}
\end{equation}
Here, $\mu_\text{B}$ is the Bohr magneton, $e$ the electron charge, $\mathbf{u}=- \frac{\mu_\text{B}}{ e M_\text{s}} P\mathbf{j}$ is the effective magnetization velocity and $P$ is the degree of spin polarization of the electron current. Under the two current model this is defined as $P=\frac{\rho_\downarrow-\rho_\uparrow }{\rho_\uparrow + \rho_\downarrow}$ \cite{Fert.1976, Haidar.2013}, where $\rho_\uparrow$ is the resistivity of the majority electrons and $\rho_\downarrow$ that of the minority electrons.

The last term in Eq.~\eqref{eq:LLGSTT} accounts for the current-induced (electric field-driven) modification to the magnetic losses\cite{ Zhang.2004, Thiaville.2005, Garate.2009}. $\beta$ is the so-called non-adiabatic STT parameter and quantifies the magnitude of such modifications. Given the formal similarity of this term with the Gilbert one, it can be considered as a current-induced component of damping \cite{Haidar.2012}.

\subsection{Current-induced spin-wave Doppler shift and spin-polarization}
To measure the spin-wave Doppler shift we can follow the current-induced modification of the phase of $\Delta L_{ij}$. Indeed, in Fig.~\ref{fig:dL} we see directly how the spectra shift in opposite directions for the two current polarities. The shift's magnitude is different for the two propagation directions since, on top of the Doppler shift, there is an additional frequency shift due to the current-induced Oersted field. These two can be separated by a symmetry analysis \cite{Supplemental, Haidar.2014}. Note that there is a clear amplitude non-reciprocity between the spectra of the two propagation directions. This is expected in the Damon-Eshbach geometry and does not affect the frequency shifts \cite{Gladii.2016}.

\begin{figure}[ht]
\includegraphics[width=0.45\textwidth]{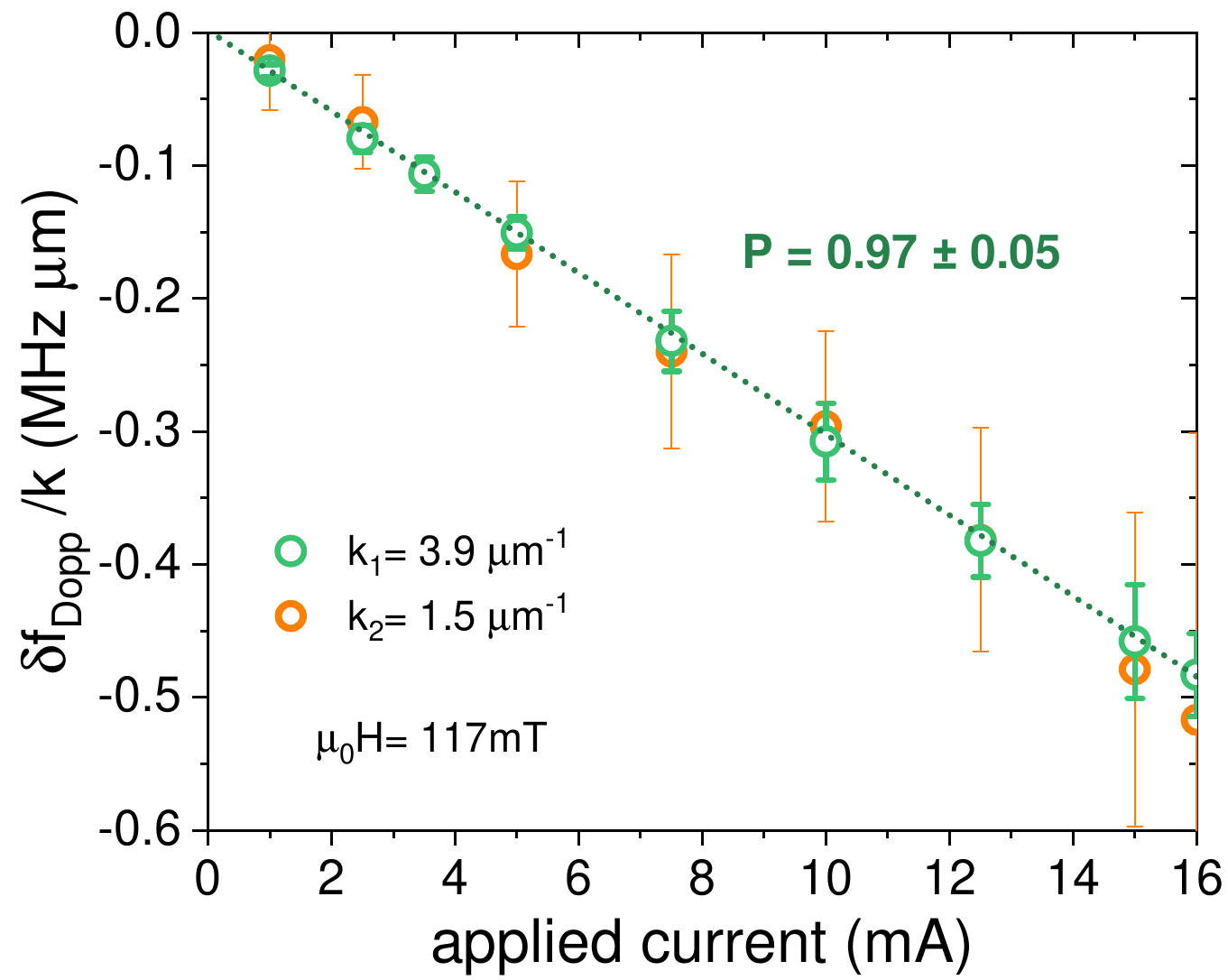}%
\caption{\label{fig:Dopp} Extracted spin wave Doppler shift as function of the applied electric current for wave vectors $k_1$ and $k_2$ at applied magnetic field $\mu_0H=117$ mT for device WA22 in Table I in \cite{Supplemental}. The dotted line is the linear fit of data corresponding to $k_1$ to Eq.~\eqref{eq:DopplerShift} from which we obtain an exemplary $P=0.97\pm0.05$.}
\end{figure}

In Fig.~\ref{fig:Dopp} we plot the extracted Doppler shift as a function of the applied electric current for the spin waves with the two experimentally accessible wave vectors $k_1$ and $k_2$ for one of the fabricated devices \cite{Supplemental}. We observe a linear dependence with very similar slope for both wave vectors ($\pm$5$\%$). The main difference between the two data sets comes down to the error bars. This difference originates from the lower signal-to-noise ratio for the $k_2$ peak compared to the $k_1$ one. By fitting these data sets to Eq.~\ref{eq:DopplerShift} we obtain the slope and having measured $\mu_0M_\text{s}=1.33 \pm 0.02$ T \cite{Supplemental}, we determine an average value for the spin polarization of $P=0.99 \pm0.08$. This value is obtained as the average between the values of $P$ obtained for the two measurable wave vectors, different magnetic fields and 5 different devices \cite{Supplemental}. The relatively small uncertainty speaks of the reproducibility of our measurements. 

In result, we measure a 100$\%$ degree of spin-polarization confirming that the diffusive electric current has also a half-metallic character in patterned Co$_2$MnSi at room temperature. Note that this is not a trivial result as previously we only had information about the film's surfaces density of states near the Fermi level~\cite{Andrieu.2016, Guillemard.2019, Melo.2021}. 

Indeed, Co$_2$MnSi/MgO/Co$_2$MnSi magnetic tunnel junctions measurements have shown that minority states contribute to the tunneling current \cite{Ishikawa.2009, Ishikawa.2009b}. Up to this point it was not clear if such interface states could contribute to the bulk diffusive transport of Co$_2$MnSi~\cite{Mott.1964}. Notably, our result shows that the electric current has a half-metallic character which means that any possible surface or bulk minority electronic state at the Fermi level does not contribute significantly to the diffusive electron transport.

We stress that our result does not imply that the density of states is fully spin polarized. Ab initio work has shown how thermal spin-fluctuations can lead to shifts in the density of states and to a slight reduction of its spin-polarization in Co$_2$MnSi \cite{Yamashita.2025}. This is not in disagreement with our result, since the properties of the density of states do not translate directly to the electric current \cite{Mazin.1999}. The current and its spin-polarization (the quantities that are relevant for most spintronic devices) are also determined by how the electron bands shift, diffuse/broaden with temperature, and how they couple to other degrees of freedom like phonons and magnons \cite{Ebert.2015, Solano.2025}. 

To conclude this part, we remark that we measure systematically a fully spin polarized electron current in several patterned devices \cite{Supplemental}. This is a true testament of the robustness of the half-metallic character of these films against nanofabrication.

\subsection{Current-induced spin-wave amplitude change}
Now that we have seen how the electric current shifts the resonance frequency of the spin waves, we turn to investigate how it modifies their attenuation.

\begin{figure}[ht]
 \includegraphics[width=0.45\textwidth]{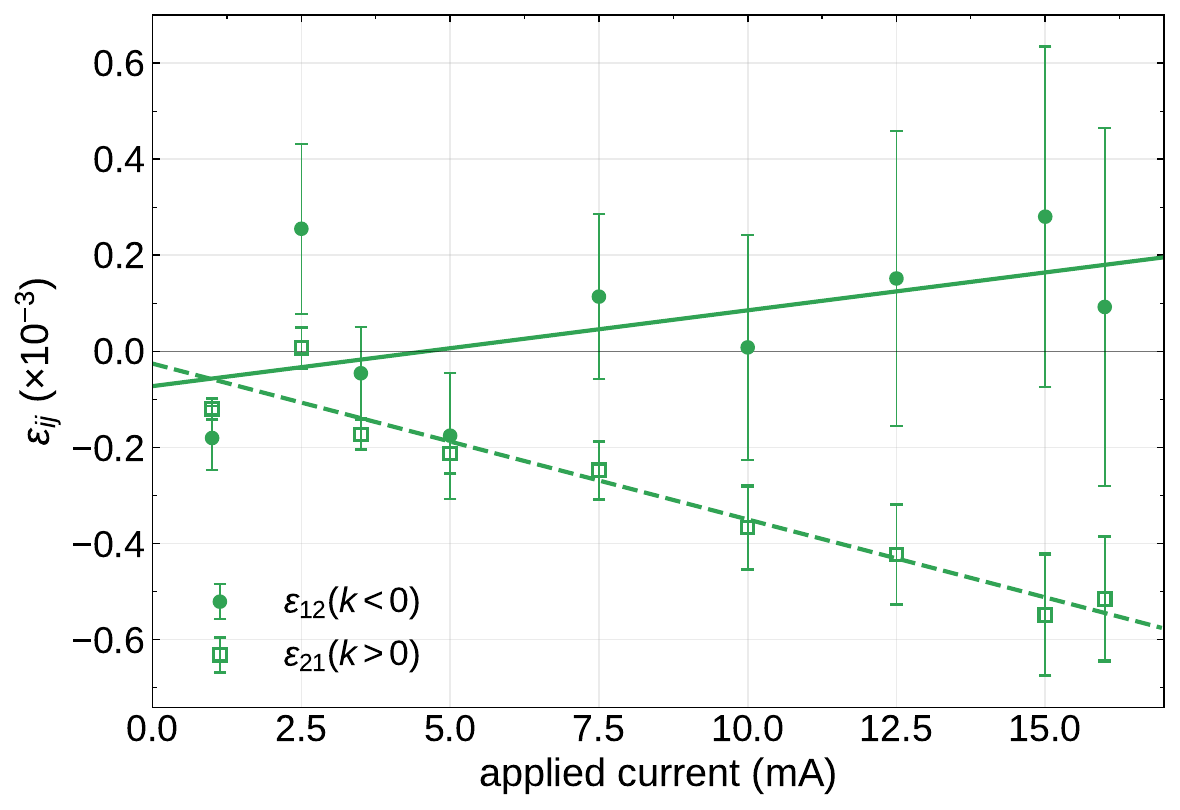}%
\caption{\label{fig:Epsilon}Current-induced modification to the mutual inductance amplitude as function of the applied current for a wave vector $k_1$= 3.90 rad/$\mu$m and and applied magnetic field $\mu_0H=117$ mT for device WA22 in Table I in \cite{Supplemental}. The lines are linear fits.}
\end{figure}

From Eq.~\eqref{eq:LLGSTT} we know that the non-adiabatic part of the STT translates into a current-induced modification of spin-wave losses. To estimate such change, we follow precisely the current-induced change of the amplitude of the propagating spin-wave signals $\Delta L_{ij}$. In the linear regime we can describe this change with:

\begin{equation}
\begin{aligned}
    \label{eq:mutualInductancesSTT}
   \Delta L_{ij}(f, I)= (1 \pm \varepsilon_{ij}(I)) g_{ij}(f+\delta\omega_{Dop}),
\end{aligned}
\end{equation}

where $g(f)$ is a waveform function (in practice we use a complex Gaussian function) and the $\varepsilon_{ij}(I)$ parameterize the current-induced amplitude change of the signal~\cite{Supplemental}. For our measurements, the factor $1+\varepsilon_{ij}=e^{-T_2/t_p}$ ($T_2$ is the characteristic relaxation time given by the Gilbert damping \cite{Supplemental,Solano.2024}) accounts for this change during the apparent time delay $t_p=1/f_p$ that it takes spin waves to travel between the two antennas ($f_p$ being the period of the oscillation in Fig.~\ref{fig:dL}, see also \cite{Supplemental}). We can extract the $\varepsilon_{ij}$ from our measured spectra \cite{Supplemental} and plot them as functions of the applied current for the two propagation directions. In Fig.~\ref{fig:Epsilon} we present an example of this data showing a linear behavior for both propagation directions. Despite the larger data dispersion for $k<0$ due to the smaller signal to noise ratio, the overall trend of the $\varepsilon_{ij}$ changes sign depending on the relative direction of spin-wave propagation and applied current. This indicates that the current-induced attenuation can either increase or decrease the net dissipation, in agreement with the expected effect of non-adiabatic STT (see Eqs.~\eqref{eq:LLGSTT}--(\ref{eq:DopplerShift})). Similar to the first part of the analysis, we use the two propagation directions and two current polarities to eliminate spurious contributions that do not follow the non-adiabatic STT symmetry. By fitting the data of several devices to the corresponding linear expressions \cite{Supplemental}, and extracting the slopes, we obtain an average current-induced modification to the damping parameter $\alpha-\beta=-0.04\pm0.01$. This is obtained as the weighted average between the values of $\alpha-\beta$ estimated for different devices and fields~\cite{Supplemental}.

To further understand this observation, we measured the temperature dependence of the magnetic damping in a piece of material that underwent nanofabrication. Our ferromagnetic resonance measurements  show that $\alpha$ follows a conductivity-like behavior as a function of the temperature \cite{Supplemental}. According to the Kambersky's torque model \cite{Kambersky.1976}, this means that $\alpha$ is dominated by intra-band electron transitions, i.e. magnetization precession leads to fluctuations of the electron states occupation around the Fermi level (breathing Fermi surface model)~\cite{Garate.2009}. In the present case of our 100$\%$ spin polarized Co$_2$MnSi, only majority electrons should contribute. 

Now, if we consider our estimated $\alpha=1.2\pm0.1\times10^{-3}$\cite{Melo.2021, Supplemental}, we obtain $\beta=0.04\pm0.01$ at room temperature \footnote{Note that despite the large $\beta$, due to the small magnitude of $\delta f_\text{Dopp}$ the maximum modification to the damping amounts to $\alpha_\text{eff}(\pm I)=\alpha \pm\beta\frac{\delta f_\text{Dopp}}{f_\text{res}}=1.2\times10^{-3}\pm 4\times10^{-6}$ at maximal current}. This means that although the magnetic damping in our material is the smallest of all metallic ferromagnets \cite{Andrieu.2016, Melo.2021}, the non-adiabatic STT remains relatively large and comparable to others (e.g Permalloy)~\cite{Sekiguchi.2012, Chauleau.2014}. 

This result becomes more surprising if one considers the historical development of STT around domain-wall motion \cite{Thiaville.2005} and measurements in $\text{Ni}_{80}\text{Fe}_{20}$ where the estimated ratio $\beta/\alpha$ is close to unity \cite{Bailleul.2017, Chauleau.2014}. Actually, under the Kambersky torque model, both $\alpha$ and $\beta$ originate from the same electron transitions \cite{Garate.2009}. Yet, they differ in that each electron state contributes to $\beta$ with a weight $v_n/u$ proportional to its respective band velocity $v_n$ [$n$ is the electron band index and $u$ is the magnetization velocity, see Eq.~\eqref{eq:DopplerShift}]~\cite{Garate.2009}. In a hypothetical fully Galilean invariant system (all electrons have the same velocity and $v_n/u=1$) one would expect $\beta/\alpha=1$ strictly. However, in real materials, while for $\alpha$ all electronic transitions contribute with the same weight, for $\beta$, electron states with large velocities would have the main contributions (i.e. $\beta/\alpha\neq1$). Therefore, we argue that the ratio $\beta/\alpha$ could be interpreted as a probe of the type of electron states involved in both dissipative processes. 

Indeed, the proven half-metallicity of the transport in our films (reduced type and number of electron transitions allowed) and the small value of $\alpha$ allow us to interpret the estimated ratio $\beta/\alpha\approx40$ closely to Kambersky's model: in Co$_2$MnSi, the large $\beta$ is an experimental signature that fast majority electron states around the Fermi level (highly conductive electrons) do in fact contribute to the magnetic dissipation.

\section{\label{sec:Conclusion}
Conclusions}

We have had a closer look at the nature of the electronic states involved in the spin transfer torque processes between an electron current and propagating spin waves in the Heusler compound Co$_2$MnSi.

Our study of the current-induced spin-wave frequency shifts reveals that the diffusive electric current is fully spin-polarized, i.e. it contains only majority electron states, in agreement with the predicted half-metallic character of the material. Remarkably, this occurs in patterned devices of thin films at room temperature, which entails a promising realistic functionality for spintronic applications. 

Additionally, we measured the current-induced change of the mutual inductance amplitude. From this we estimated a relatively large non-adiabatic spin transfer torque parameter which is attributed to the contribution of highly conductive majority states to intra-electron-band magnetic dissipation processes. We think these experimental observables can assist in the study of electronic transitions and their contributions to magnetic losses particularly in half metals and highly spin-polarized-low-damping metallic ferromagnets.

\begin{acknowledgments}
We thank Marc Lenertz for the X-ray characterization; Romain Bernard, Sabine Siegwald and Hicham Majjad for technical support during nanofabrication at STnano platform, Arnaud Boulard, Benoît Leconte, Daniel Spor, Jérémy Thoraval and Fares Abiza for their support in assembling and testing the broadband ferromagnetic resonance and propagating spin-wave spectroscopy setup; Sylvie Migot for the focus ion beam preparation and Jaafar Ghanbaja for transmission electron microscopy measurements; Shogo Yamashita for theoretical calculations and discussions. We acknowledge financial support from the Interdisciplinary Thematic Institute QMat, as part of the ITI 2021-2028 program of the University of Strasbourg, CNRS and Inserm, IdEx Unistra (ANR 10 IDEX 0002), SFRI STRAT’US project (ANR 20 SFRI 0012) and ANR-17-EURE-0024 under the framework of the French Investments for the Future Program. We also acknowledge financial support from Region Grand Est through its FRCR call (NanoTeraHertz and RaNGE projects) and from Agence Nationale de la Recherche under contract No. ANR-20-CE24-0012 (MARIN), ANR-22-EXSP-0004 (SWING, France 2030 PEPR Spin program) and ANR-20-CE24-0023 (CONTRABASS). Anna Maria Friedel acknowledges support from the Franco-German University (FGU).
\end{acknowledgments}
\bibliography{HeuslerP}
\bibliographystyle{apsrev4-2}
\end{document}


\preprint{APS/123-QED}
\title{Supplemental material: \\
Spin-polarization of the electric current in half-metallic Co$_2$MnSi Heusler thin films}

\maketitle

\section{\label{sec:SampleGrowth}Sample growth and \\ structural characterization\protect}
In this section we present a summary of additional details the reader of the main manuscript may find interesting about the films' growth, structural and chemical characterization. For even further details refer to the thesis of A. Friedel \cite{Friedel.2025}.
\subsection{Growth}
The films are grown on a 30 x 30 mm$^2$ MgO commercial substrate (lattice mismatch of 5$\%$ with bulk Co$_2$MnSi). The molecular beam epitaxy was performed on a COMPACT 21 EB 200 MBE system from RIBER. First, a 3x3 cm$^2$ MgO(001) substrate was outgassed and then covered with a 10 nm MgO  buffer layer at a substrate temperature of 900°C. The sample was then left to cool down, while the flux densities of Co, Mn and Si were calibrated for the co-deposition step. Next, the nominally 20 nm Co$_2$MnSi Heusler layer was grown by co-deposition of Co, Mn and Si at 400°C. The growth procedure is described in more detail by Guillemard et al. in reference \cite{Guillemard.2020}. Subsequently, the sample was annealed at 600°C. Annealing reduces the surface roughness and improves the ratio of desired chemical L2$_1$ ordering to undesired B2 disorder in the Heusler film \cite{Gaier.2008}. Finally, a 9 nm MgO insulating layer was grown and capped with a 6.7 nm Ti film. 

The good cristallinity of the layers was monitored during growth by reflective high energy electron diffraction and after growth by X-Ray diffraction with a quality comparable to that reported before \cite{Guillemard.2019} . Using X-ray reflectometry we measure an actual thickness of $t$=20.0 $\pm 0.2$ nm for the Co$_2$MnSi layer, which is corroborated with TEM micrographs (see Fig.~\ref{fig:stack}).

After growth, the 30 x 30 mm$^2$ thin film sample was cut into 9 sample pieces of 10 x 10 mm$^2$ to allow for independent processing.

\subsection{X-ray diffraction}
\begin{figure}[H]
\centering
\includegraphics[width=0.9\textwidth]{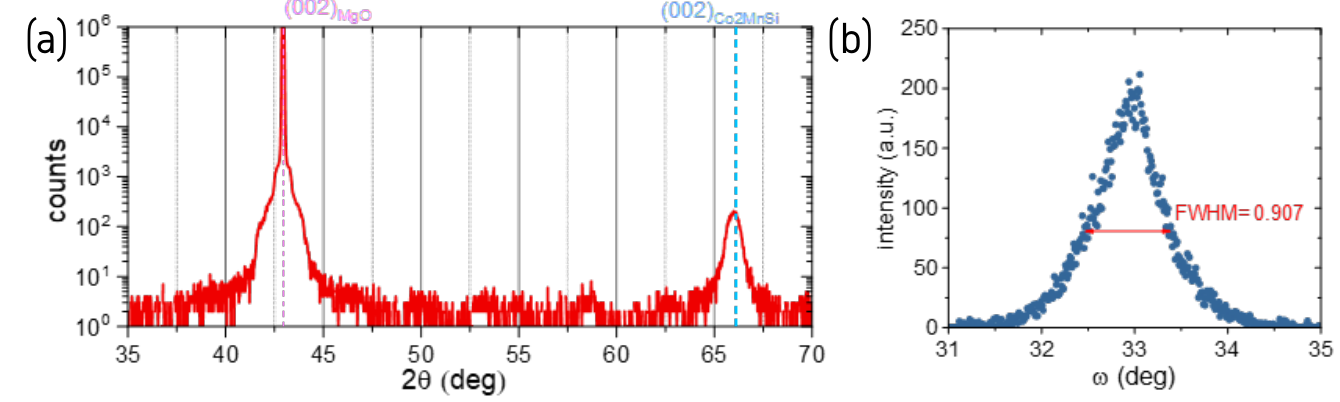}
\caption{(a) X-ray diffraction scan showing the substrate peaks and the Co$_2$MnSi's peak centered at 66.047$\degree$. The dashed lines show the expected possition for the corresponding planes of bulk MgO and Co$_2$MnSi. (b) Rocking curve measured for the Co$_2$MnSi's peak.} 
\label{fig:xray}
\end{figure}

X-ray diffraction and rocking curve measurements [Fig.~\ref{fig:xray}] were performed in a Smartlab Rigaku thin film diffractometer in parallel beam geometry equipped with a 9kW copper rotating anode and a double bounce Ge(220)x2 front monochromator ($\lambda_{\text{K}{\alpha1}}$ = 0.15406 nm).

The out-of-plane lattice constant for the Co$_2$MnSi layer estimated from the diffraction measurements is $a_{[001]}$=0.55 nm. It is smaller than the expected in-plane value $a_{[001]}=0.566 nm$ \cite{Guillemard.2020}. This is possibly evidence of a tetragonal distortion due to the lattice mismatch with the substrate. The full width at half maximum obtained from the rocking curve is 0.907$\degree$. 

\subsection{Transmission electron microscopy}
\begin{figure}[H]
\centering
\includegraphics[width=0.6\textwidth]{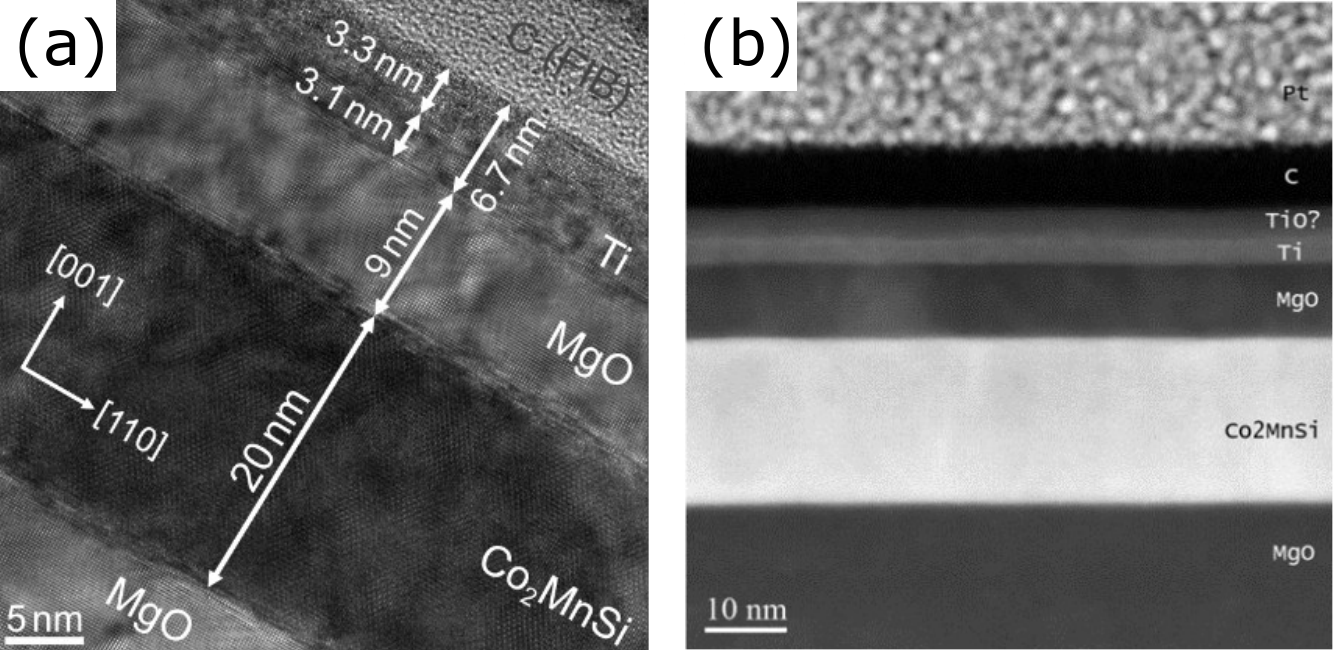}
\caption{ TEM micrographs of the deposited stack (a) High-resolution TEM image. (b) High angle annular dark field-scanning TEM microscopy image. A contrast is visible across the Ti layer.} 
\label{fig:stack}
\end{figure}

From one of the cut pieces of sample, a lamella was cut by Ga+ focused ion beam. The ion beam etching was performed along the [110] direction of the Co$_2$MnSi lattice. To cover the region of interest carbon was deposited to improve conductivity and platinum for protection. The thickness of the resulting lamella was roughly 50nm at the TEM/STEM probing area (thin film cross-section). 
Transmission electron microscopy (TEM) experiments were carried out on a JEM-ARM 200F Cold FEG
TEM/STEM system operating at 200 kV. High resolution measurements revealed an homogeneous film part of which is shown in Fig.~\ref{fig:stack} (a). Scanning TEM (STEM) using a high angle annular dark field (HAADF) revealed also a change of contrast across the Ti capping layer thickness [Fig.~\ref{fig:stack} (b)]. We attribute this to an oxidation of this layer after being exposed to air.

\begin{figure}[H]
\centering
\includegraphics[width=0.8\textwidth]{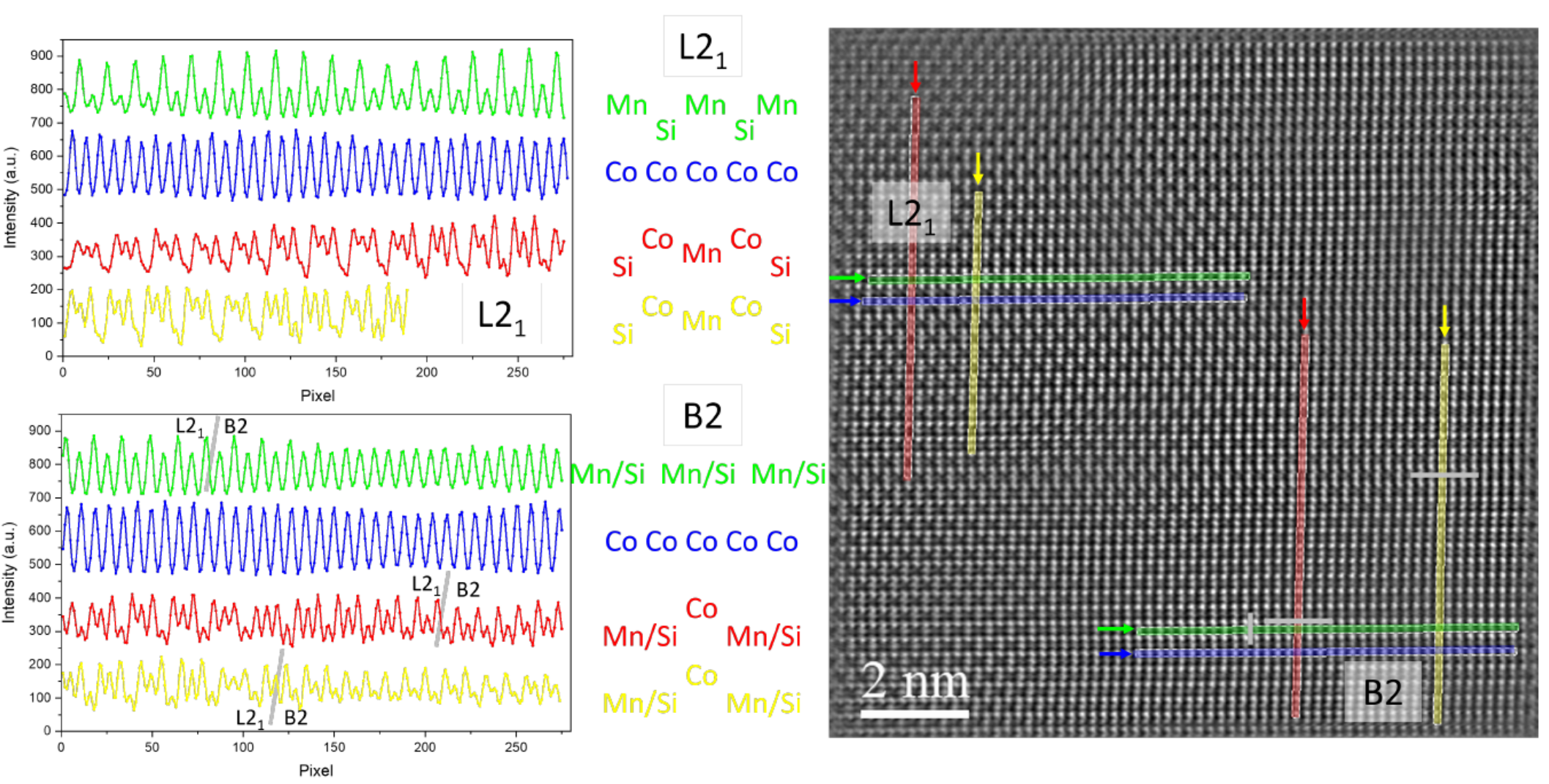}
\caption{Intensity profiles (left) extracted from a Fourier filtered HAADF-STEM micrograph (right). The gray marks in the adjacent intensity pattern (bottom left) indicate where a change of L21 order to B2 disorder is observed. The corresponding position in the HAADF-STEM micrograph is indicated by gray lines perpendicular to the green, red and yellow marked areas from which the integrated intensity was extracted.} 
\label{fig:chemicalAnalysis}
\end{figure}

Additionally, from the HAADF-STEM measurements we performed an intensity pattern analysis which can be related to the chemical ordering of the alloy~\cite{Guillemard.2020}. In Fig.~\ref{fig:chemicalAnalysis} we present both an example of a Fourier filtered HAADF-STEM micrograph and its respective intensity patterns. Performing such an analysis on various HAADF-STEM micrographs taken at different sites across the 7 µm long lamella, allowed us to conclude that the Co$_2$MnSi layer is predominantly formed by the perfectly ordered chemical structure L2$_1$ with only a small fraction of the disordered structure B2 \cite{Guillemard.2020}. 

\section{\label{sec:characterization}Magnetic characterization\protect}
In this section we present a summary of additional details the reader of the main manuscript may find interesting about the magnetic characterization of the films. For further details please refer to the thesis of J. Solano \cite{Solano.2024}.

\subsection{SQUID magnetometry}

To measure the saturation magnetization of the films we used a Quantum Design MPMS3 SQUID magnetometer. One of the 10x10 $\text{mm}^2$ pieces of film was cut into 2x2 $\text{mm}^2$ pieces and then glued to a quartz holder with transparent nail polish. During the measurements, the magnetic field $H$ is applied parallel to the film's plane and along Co$_2$MnSi's [110] easy axis.

To accurately determine the magnetic moment $\mu(H)$, we used a protocol developed by Amorin et al.\cite{Amorim.2021} that provides a correction factor $\alpha'$ for the measurement of the magnetic moment. This correction factor is determined from the measurements themselves by exploiting the relationship between the moment values determined from the DC and VSM modes of the magnetometer. Post measurement, diamagnetic contributions were corrected by determining the diamagnetic susceptibility at high fields and subtracting it from the hysteresis loops. Finally, we used the film thickness $t$ (measured by X-ray reflection and TEM) and the measurement of the surface area $S$ of the films by optical microscopy to estimate the magnetization $M(H)= \frac{1}{\alpha'} \frac{\mu(H)}{t S}$.
 
\begin{figure}[H]
\centering
\includegraphics[width=0.45\textwidth]{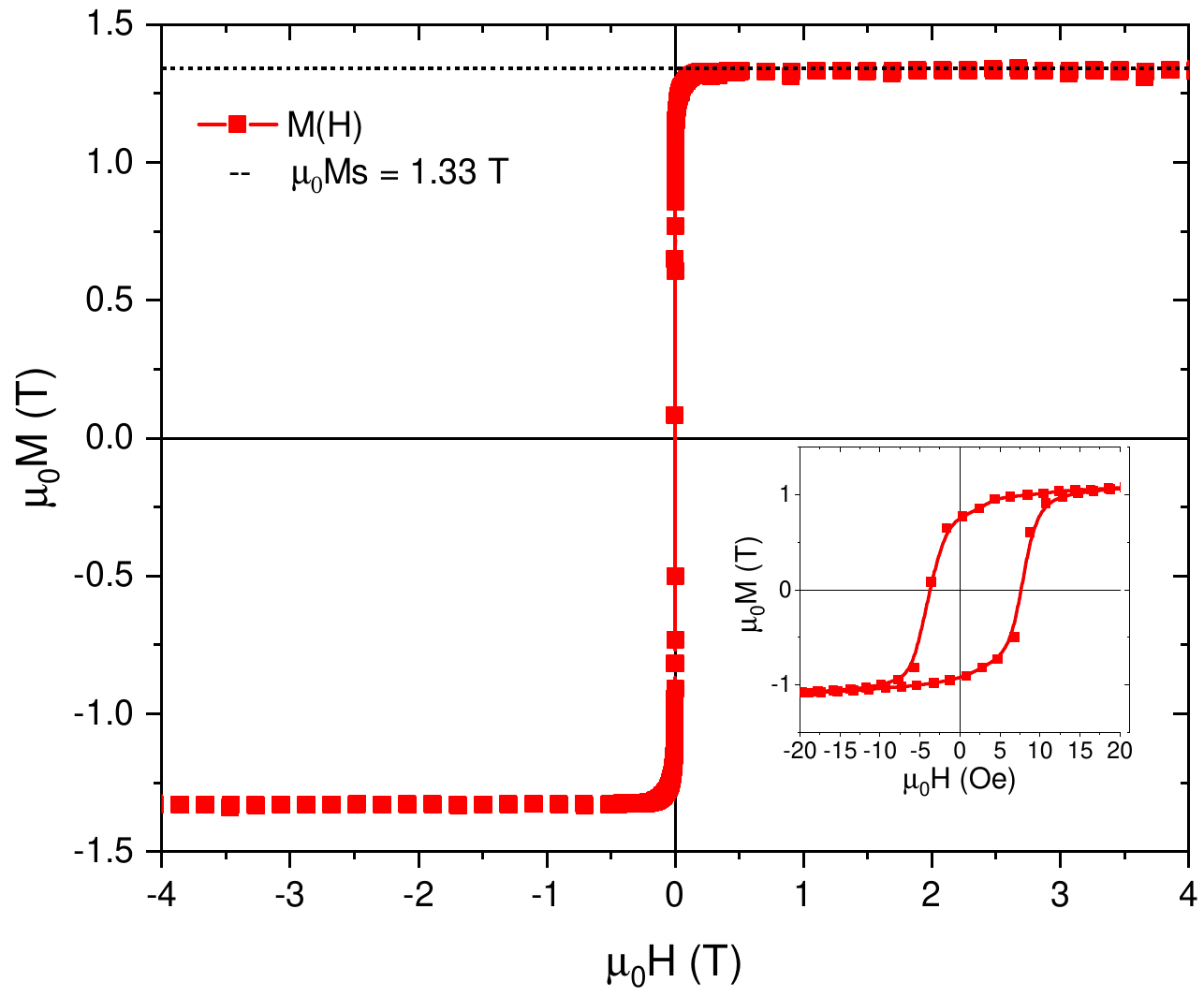}
\caption{Magnetization hysteresis loop where the external magnetic field is applied along Co$_2$MnSi's [110] direction. The inset shows a magnification of the loop.} 
\label{fig:MvsT}
\end{figure}

\subsection{Ferromagnetic resonance}

We conducted our ferromagnetic resonance characterization using a protocol similar to the one presented in ref.~\cite{Solano.2022}. The main difference is that we inserted our sample (2$\times$2 mm$^2$ film piece) and measuring PCB in a cryostat to vary the temperature. Unfortunately, this does not allow us to fully calibrate the microwave network as a function of the temperature forcing us to proceed in a semi-calibrated state: we measure precisely the electrical delays from each port to the sample as a function of temperature and we correct them manually in their respective VNA port extensions. We cannot calibrate the losses in the network and therefore neither the amplitude of the signal. However, a large part of extrinsic effects is taken care by the subtraction of background using a reference magnetic state at large fields (typically 2.2 T). In this way the ferromagnetic resonance is resolved without major extrinsic effects from the microwave network. 

\begin{figure}[H]
\centering
\includegraphics[width=0.85\textwidth]{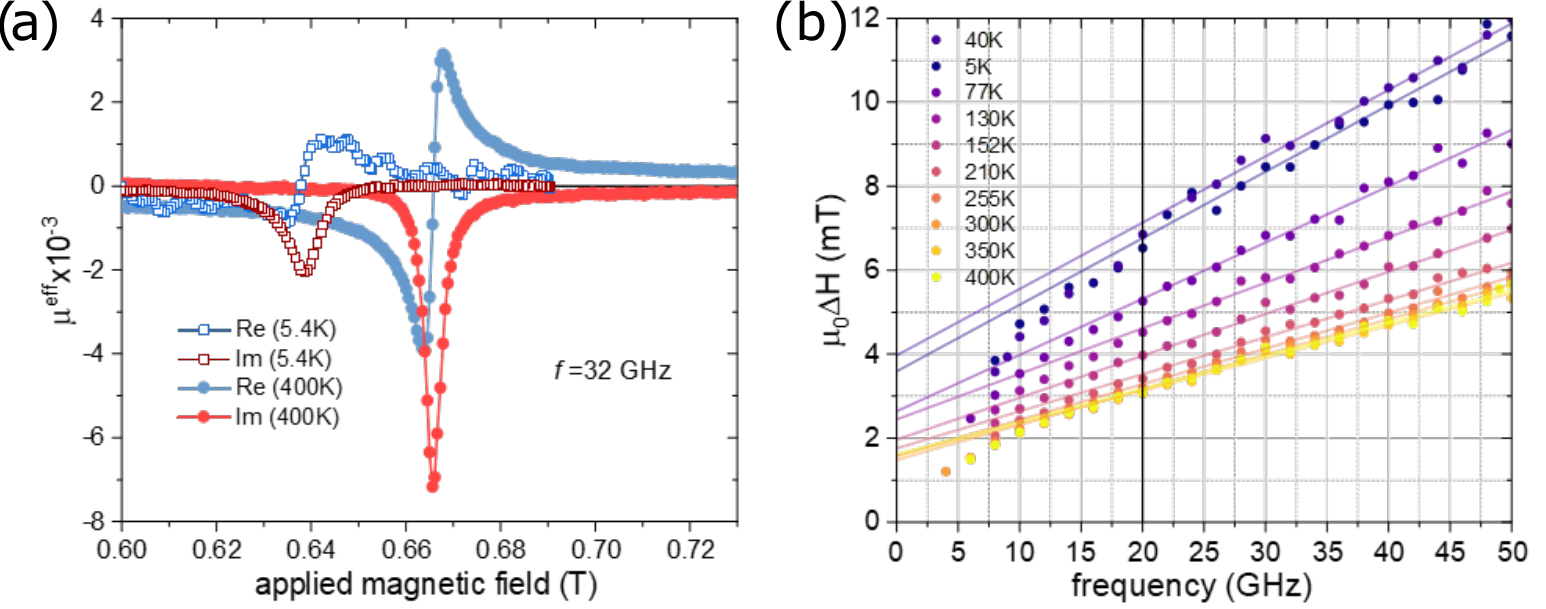}
\caption{(a) Ferromagnetic resonance for a thin film Co$_2$MnSi (20 nm) measured at different temperatures for in-plane applied magnetic field (along Co$_2$MnSi's [110] direction). (a) Real and imaginary part of the ferromagnetic resonance spectra (FMR) at 32 GHz for 5.4 K and 400 K. (b) FMR linewidth at half maximum as function of the frequency for different temperatures. The lines are the corresponding linear fits from 20GHz to 50GHz.} 
\label{fig:FMR}
\end{figure}

In the measurements we fix the microwave excitation at a given frequency (nominally at a power of -15 dBm $\approx$ 32 µW), and sweep the applied magnetic field around the resonance condition. We apply the magnetic field in-plane (along Co$_2$MnSi's [110] direction). Despite the not fully calibrated state of the network, this method allow us to follow clearly the ferromagnetic resonance as a function of the applied magnetic field [Fig.~\ref{fig:FMR}].

By fitting the resonance peaks [Fig.~\ref{fig:FMR} (a)] to a Lorentz complex function, we obtain the resonance field and the linewidth at half maximum for each frequency as a function of the temperature. We plot the linewidth as a function of the frequency and for different temperatures in Fig.~\ref{fig:FMR} (b). 

For each temperature, we fit the linewidth data to the linear relationship between linewidth ($\Delta H$) and frequency:

\begin{equation}
\begin{aligned}
    \label{eq:linewidth}
    \Delta H = 2\frac{2\pi\alpha_{FMR}}{\gamma \mu_0} f + \Delta H_o ,
\end{aligned}
\end{equation}

Here $\gamma/(2\pi)$=28.25 GHz/T is the gyromagnetic ratio, $\mu_0$ is the vacuum permeability, $\alpha_{FMR}$ is the effective magnetic damping and $\Delta H_o$ the inhomogeneous broadening. We restrict the fits down to 20 GHz as below this frequency we observe a non-linear behavior that is probably related to 2-magnon scattering as it is not present when the magnetic field is applied along the out-of-plane direction of the film~\cite{Solano.2024}. 

We remark that we observe the same magnetic damping before and after nanofabrication. Furthermore, this is in agreement with previous experimental results \cite{Melo.2021}. This is an additional indicator of the robustness of the films against fabrication processes.

\subsection{Temperature dependence of the damping and inhomogeneous broadening}

\begin{figure}[H]
\centering
\includegraphics[width=0.85\textwidth]{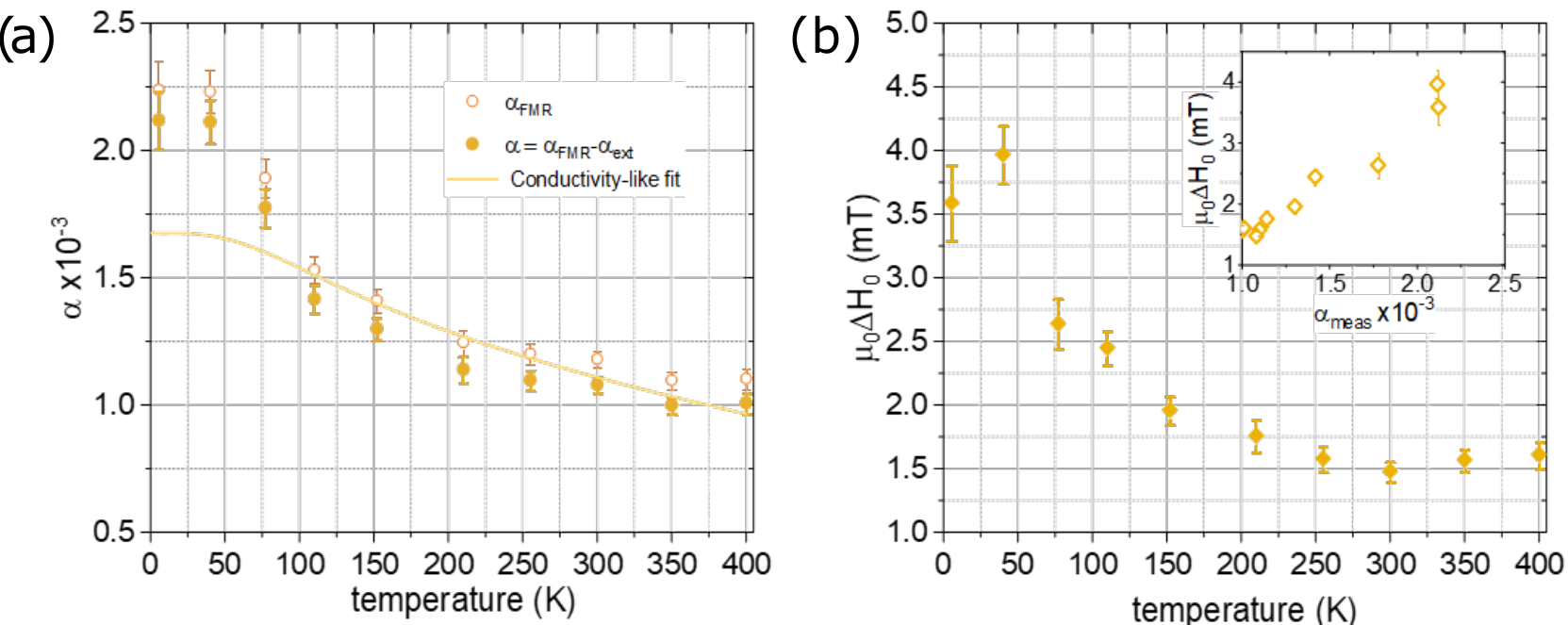}
\caption{(a) Effective $\alpha_{FMR}$ and intrinsic $\alpha$ magnetic damping as functions of the temperature. The line is the fit to the films conductivity-like function (renormalized conductivity vs. temperature). (b) Inhomogeneous linewidth as a function of the temperature. Inset: inhomogeneous linewidth as a function of the intrinsic magnetic damping.} 
\label{fig:Damping}
\end{figure}

From the fits in Fig.~\ref{fig:FMR} (b) we obtain the temperature dependence of both $\alpha_{FMR}$ and $\Delta H_o$. For the magnetic damping, we use models to account for extrinsic contributions $\alpha_{ext}$ due to the radiative damping and the damping created by eddy currents \cite{Solano.2024}. We use our measurement of the film's resistivity vs. temperature to calculate the conductivity and to fit the intrinsic damping to it in Fig.~\ref{fig:Damping} (a). The result is a conductivity-like temperature dependence only, which captures most of the experimental temperature dependence. We think the disagreement with the conductivity-like model [Fig.~\ref{fig:Damping} (a) for T < 100 K] could originate from an  overestimation of the experimentally-extracted damping at low temperatures due to two-magnon scattering effects and a correlation with the inhomogeneous broadening which can be observed in the inset of Fig.~\ref{fig:Damping} (b). Nevertheless, we argue that most of the damping can be modeled by a conductivity-like dependence since the data does not admit a resistivity-like component. Therefore, intra-band electron transitions should be the main contribution to the magnetic dissipation \cite{Garate.2009, Kambersky.1976}. This is in agreement with what one would expect for a half-metal since intra-band (no-spin-flip) electronic transitions are expected to be predominant, and inter-band (between spin bands requiring spin-flip) expected to be heavily suppressed~\cite{Garate.2009, Kambersky.1976}.

\section{Propagating spin wave spectroscopy}
In this section we present a summary of additional details the reader of the main manuscript may find interesting about spin wave Doppler experiments. For even further details, refer to the thesis of J. Solano \cite{Solano.2024}.

\begin{figure}[H]
\centering
\includegraphics[width=0.70\textwidth]{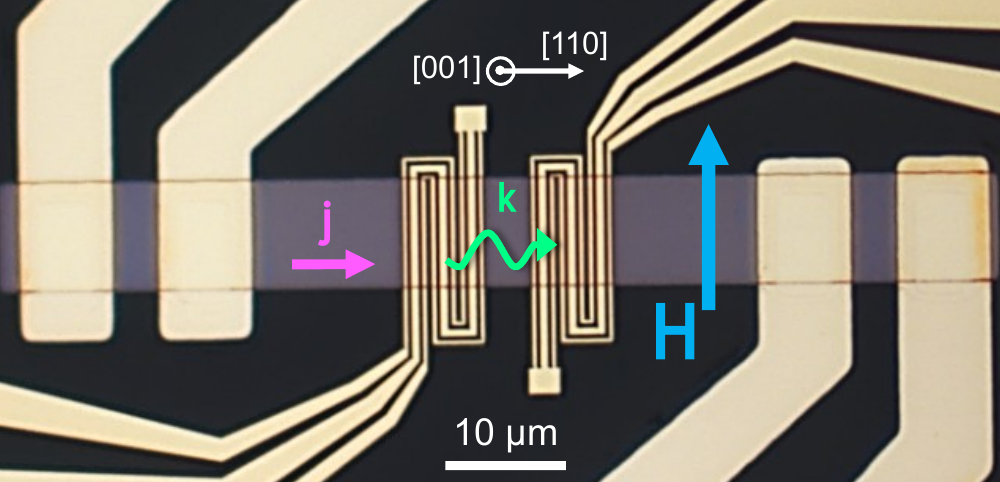}
\caption{Microscope picture of spin wave Doppler device with a sketch of the experimental geometry. The ferromagnetic strip is the horizontal bar at the center, the 4 lateral leads ar used to perform 4-probe DC measurements and the central structures are the microwave antennas with the Fourier transform plotted in [Fig.~\ref{fig:FTT} (a)].} 
\label{fig:Device_sup}
\end{figure}

\begin{figure}[H]
\centering
\includegraphics[width=0.90\textwidth]{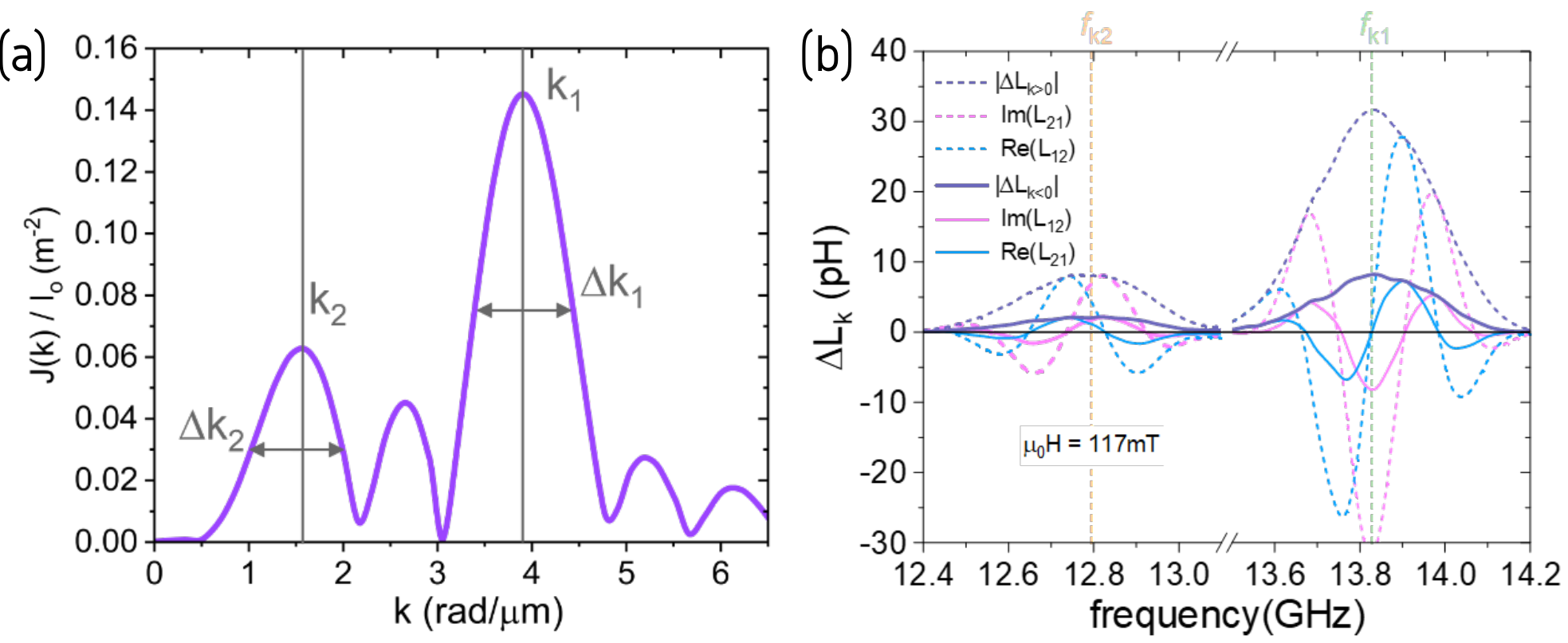}
\caption{(a) Calculation of the Fourier transform of the current density of an antenna displaying maxima around the wave vectors $k_1$ = 3.9 rad/$\mu$m and $k_2$ = 1.6 rad/$\mu$m assuming a uniform current density in the strands of the antennas. (b) Measured mutual inductance $\Delta L$ between the antennas as a function the frequency for an applied field $\mu_0 H$=117 mT. Solid lines correspond to $k<0$, and dashed lines to $k>0$. The edge-to-edge distance between the antennas in this case is 3 $\mu$m and the Co$_2$MnSi strip has a width of $w$=15 $\mu$m.} 
\label{fig:FTT}
\end{figure}

Measurements are nominally performed at a source power of -15 dBm (0.03 mW) or -10 dBm (0.1 mW). The geometry of our antennas (see Fig.~\ref{fig:Device_sup}) produces two main excitation peaks around the wave numbers $k_1$=3.9 rad/$\mu$m and $k_2$=1.6 rad/$\mu$m [Fig.~\ref{fig:FTT} (a)], as we can see $k_2$ has a less efficient excitation and therefore a lower amplitude to noise ratio. In Fig.~\ref{fig:FTT} (b) we can observe the corresponding measured spin wave spectra. To describe this change in mutual inductance signal ($\Delta L$), we use a complex Gaussian function~\cite{Haidar.2012}:

\begin{equation}
\begin{aligned}
    \label{eq:mutualInductance}
   \Delta L(f, \mathbf{k}) = \Delta L^0_{ij}(f)= \frac{A}{w\sqrt{\pi/2}} e^{ -\frac{2(f-f_k)^2}{w^2} } e^{ i \frac{ 2\pi(f-f_{r})}{f_{p}} },
\end{aligned}
\end{equation}

here $\mathbf{k}$ is the wave vector that in our experiments is always either parallel or antiparallel to the [110] crystal axis of Co$_2$MnSi (see Fig.~\ref{fig:Device_sup}). Therefore, experimentally we define the indices $ij$=21 for $k>0$ (spin wave measured at port 2 after being excited at port 1) and $ij$=12 for $k<0$ (spin wave measured at port 1 after being excited at port 2). Additionally: $f_k$ is the central frequency of the Gaussian wave packet and the resonance frequency of the spin waves, $f_p$ corresponds to the oscillation period or $\tau = 1/f_p$ is the delay time between the excitation and the detection, $w$ is the width of the Gaussian wave packet and $f_r$ is a phase-reference frequency at which the real part is maximum and the imaginary zero.

When we inject a DC electric current density $\mathbf{j}$ (with a current amplitude $I$) into the ferromagnetic strip, there is a spin transfer torque (STT) between the spin-polqrized current and the spin wave with wave vector $\mathbf{k}$. Similarly to the wave vector, the current is also applied experimentally only along the direction of the Co$_2$MnSi strip: we denote by $+I$ the case when $\mathbf{j}$ is parallel to [110] ($I>0$), and by $-I$ when $\mathbf{j}$ is antiparallel to [110] ($I<0$). 
We can include the current-induced modifications to the mutual inductance signal as follows~\cite{Haidar.2012}:
\begin{equation}
\begin{aligned}
    \label{eq:mutualInductancesSTT}
    \Delta L_{ij}(f,\pm I)= (1 \pm \varepsilon_{ij}) \Delta L^0_{ij}(f \mp \delta f_{ij})= (1 \pm \varepsilon_{ij})\frac{A}{w\sqrt{\pi/2}} e^{ -\frac{2(f-f_k \mp \delta f_{ij} )^2}{w^2} } e^{ i \frac{ 2\pi(f \mp \delta f_{ij}-f_{r})}{f_{p}} }.
\end{aligned}
\end{equation}

In this way, we define $\delta f_{ij}$ as a current-induced frequency shift (e.g. Doppler shift, but not exclusively) and $\varepsilon_{ij}$ as a current-induced relative change of amplitude (e.g. as the one produced by the non-adiabatic STT) when $I>0$. Note that both $\delta f_{ij}(k,I)$ and $\varepsilon_{ij}(k,I)$ are functions of the spin wave wave number $k$ and the current magnitude $I$, which are omitted in equations \eqref{eq:mutualInductancesSTT}, \eqref{eq:Rij} and \eqref{eq:RealImagRij} for simplicity. Furthermore, we have assumed that $\delta f_{ij}(I>0)=-\delta f_{ij}(I<0)$ and $\varepsilon_{ij}(I>0)=-\varepsilon_{ij}(I<0)$, which is in agreement with the expected symmetry of the current-induced modifications. 

To systematically extract the current-induced parameters from the measured spectra, we can calculate the following ratio \cite{Haidar.2012}:

\begin{equation}
\begin{aligned}
    \label{eq:Rij}
   R_{ij} (I)= \frac{\Delta L_{ij}(+ I) - \Delta L_{ij}(- I)}{ \Delta L_{ij}(+ I) + \Delta L_{ij}(- I) },
\end{aligned}
\end{equation}

and obtain its real and imaginary part (at first order in $\delta f_{ij}$ and $\varepsilon_{ij}$):

\begin{subequations}
    \label{eq:RealImagRij}
     \begin{align}
     Re( R_{ij} ) \approx \varepsilon_{ij} + 4 \delta f_{ij} \frac{f-f_k}{w^2} \label{eq:ReRij}\\
     Im( R_{ij} ) \approx -2\pi \frac{ \delta f_{ij} }{f_\text{p}}.\label{eq:ImRij}
    \end{align}
\end{subequations}

\begin{figure}[H]
\centering
\includegraphics[width=0.90\textwidth]{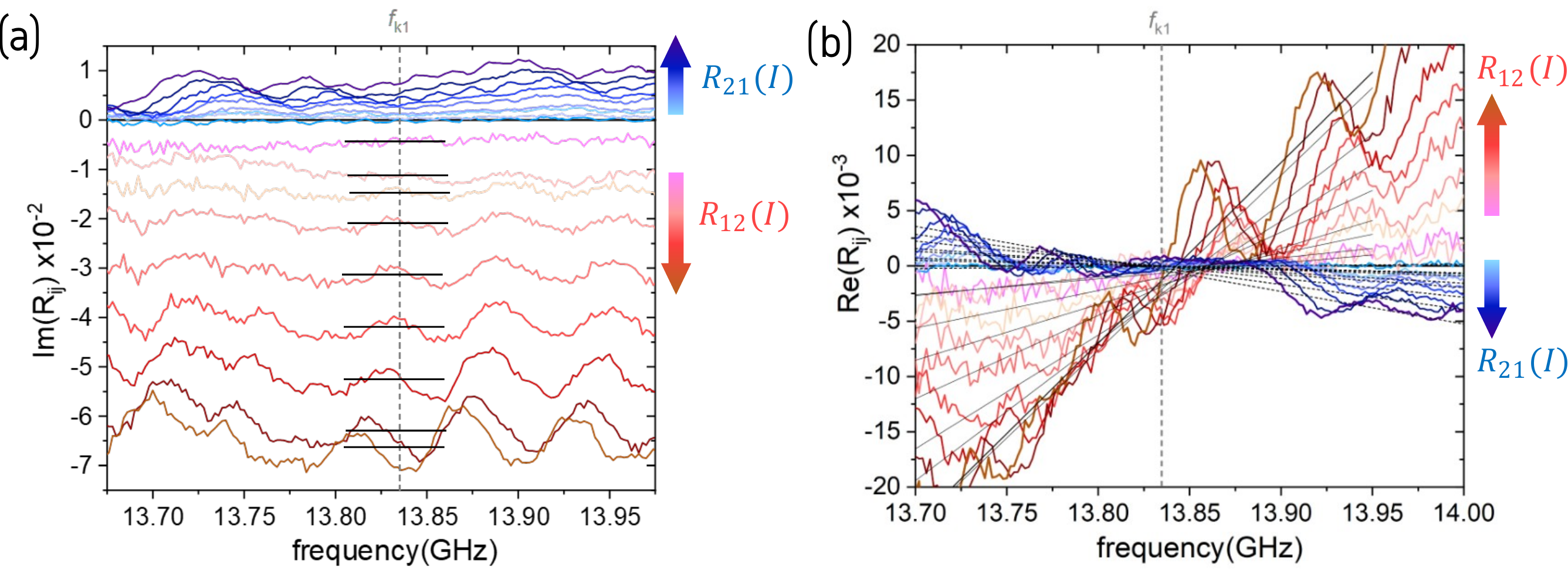}
\caption{$R_{ij}$ as a function of frequency for an applied field $\mu_0 H$=117mT and wave vector $k_1$=3.9 rad/$\mu$m for different currents: 1mA, 2.5mA, 3.5mA, 5mA, 7.5mA, 10mA, 12.5mA, 15mA, 16mA. The arrows' direction display the increasing current. (a) Imaginary parts, horizontal lines shows the fits at resonance frequency [Eq.~\eqref{eq:ImRij}]. (b) Real part, the lines are the fits to Eq.~\eqref{eq:ReRij}.} 
\label{fig:Rij}
\end{figure}

In Fig.~\ref{fig:Rij} we present the imaginary and real parts of $R_{ij}$ for $k_1$ = 3.9 rad/$\mu$m at an applied field of $\mu_0 H$=117 mT for various applied currents $|I|$ ranging from 1 mA to 16 mA. The data presents unexpected oscillations that modulate the amplitude. To obtain the current-induced frequency shifts $\delta f_{ij}$ and the current-induced amplitude changes $\varepsilon_{ij}$ we have fitted the data to Eqs.~\eqref{eq:RealImagRij} centered at $f_k$. 

\subsection{Adiabatic STT analysis: Spin wave Doppler shift\label{subsecA}}
The current-induced frequency shifts $\delta f_{ij}$ contain the spin wave Doppler shift $\delta f_\text{Dopp}$ due to the adiabatic STT and the frequency shifts due to the Oersted field produced by the injected current $\delta f_\text{Oe}$ (reciprocal), $\delta f_\text{NROe}$ (non-reciprocal)~\cite{Haidar.2012, Haidar.2014}. Then including all the frequency shifts we define:

\begin{equation}
\begin{aligned}
    \label{eq:shiftij}
    \delta f_{ij}(I) =&  \delta f_\text{Dopp}(k, I) + \delta f_\text{Oe}( I) + \delta f_\text{NROe}(k, I).
\end{aligned}
\end{equation}

For convenience we define:  

\begin{subequations}
    \label{eq:positiveK&IShifts}
     \begin{align}
     \delta f_\text{Dopp}^*(I) &= \delta f_\text{Dopp}(k>0,I>0), \label{eq:DopPositiveK&I}\\
     \delta f_\text{Oe}^*(I) &= \delta f_\text{Oe}(I>0), \label{eq:OePositiveK&I}\\
     \delta f_\text{NROe}^*(I) &= \delta f_\text{Oe}(k>0,I>0),\label{eq:NROePositiveK&I}
    \end{align}
\end{subequations}

Then for spin waves with $k>0$ we have:
\begin{equation}
\begin{aligned}
    \label{eq:shift21}
    \delta f_{21}(I) =&  \delta f_\text{Dopp}^*(I) + \delta f_\text{Oe}^*(I) + \delta f_\text{NROe}^*(I),
\end{aligned}
\end{equation}

and keeping in mind that $\delta f_\text{Dopp}$ and $\delta f_\text{NROe}$ are odd functions of $k$, while $\delta f_\text{Oe}$ is not, for $k<0$ we have:

\begin{equation}
\begin{aligned}
    \label{eq:shift12}
    \delta f_{12}(I) =& - \delta f_\text{Dopp}^*(I) + \delta f_\text{Oe}^*(I) - \delta f_\text{NROe}^*(I).
\end{aligned}
\end{equation}
    
Finally, combining the shifts of the two propagation directions to obtain \cite{Haidar.2012}

    \begin{subequations}
        \label{eq:Dopp&OerstedShiftsFromRij}
         \begin{align}
         \delta f_\text{Dopp}^*(I) &= \frac{ \delta f_{21}(I) - \delta f_{12}(I) }{2} - \delta f_\text{NROe}^*(I), \label{eq:DoppFromRij}\\
         \delta f_\text{Oe}^* (I) &= \frac{ \delta f_{21}(I) + \delta f_{12}(I) }{2}. \label{eq:OeFromRij}
        \end{align}
    \end{subequations}
An example of the current-induced spin wave Doppler shift $\delta f_\text{Dopp}$ resulting from this symmetry analysis is shown in in Fig. 3 of the main manuscript. In our estimation of $\delta f_\text{Dopp}$, we have safely neglected $\delta f_\text{NROe}^*(I)$ since its magnitude remains very small ($|\delta f_\text{NROe}^*| < 4~\text{kHz}$) for this material and experimental conditions \cite{Haidar.2014}.

To test the reproducibility of our results, we performed our Doppler measurements and determination of $P$ for devices with combinations of: 2 different strip widths $w_\text{s}$, 2 different edge-to-edge distances between the antennas $D$, 2 different applied magnetic fields $H$ and for 2 different wave vectors $k_1\text{=3.9} \; \text{rad}/\mu\text{m}$ and $k_2\text{=1.5} \; \text{rad}/\mu\text{m}$. The summary of these results can be found in Table~\ref{tab:PValuesHeusler}.

\begin{table}[H]
    \centering 
    \caption[Degree of spin-polarization of the current $P$ in Co$_2$MnSi]{ Estimated degree of spin-polarization $P$ in the Co$_2$MnSi film using spin waves with  wave vectors $k_1\text{=3.9} \; \text{rad}/\mu\text{m}$ and $k_2\text{=1.5} \; \text{rad}/\mu\text{m}$. The values are reported for different devices under different applied magnetic fields $H$. The film thickness is $20$ nm. $D$ is the distance between the internal edges of the antennas. $w_\text{s}$ is the nominal width of the ferromagnetic strip. \label{tab:PValuesHeusler} }
    \begin{tabular}{| c | c | c | c | c | c | c | c |}
    \hline
    device & $w_\text{s}\pm 0.07$($\mu \text{m}$) & $D$($\mu \text{m}$) & $\mu_0 H$(mT)  & $\rho\pm$0.5($\mu \Omega$ cm)  &  $P(k_1)$ & $P(k_2)$ \\
    \hline \hline
    WA22 & 14.14 & 3 & 117 & 37.7 & 0.97$\pm$0.05 & 1.1$\pm$0.1\\
    WA23 & 9.10 & 3 & 117 & 37.6 &  1.15$\pm$0.06 & 1.1$\pm$0.08\\
    WA23 & 9.10 & 3 & 167 & 37.6 &  1.02$\pm$0.05 & 0.90$\pm$0.05\\
    WA33 & 9.10 & 4 & 117 & 38.1 &  1.06$\pm$0.05 & 1.04$\pm$0.06\\
    WA33 &9.10 & 4 & -117 & 38.1 &  1.10$\pm$0.06 & 0.9$\pm$0.1\\
    \hline
    \end{tabular}
    \\[10pt]
\end{table}

\begin{figure}[H]
\centering
\includegraphics[width=0.80\textwidth]{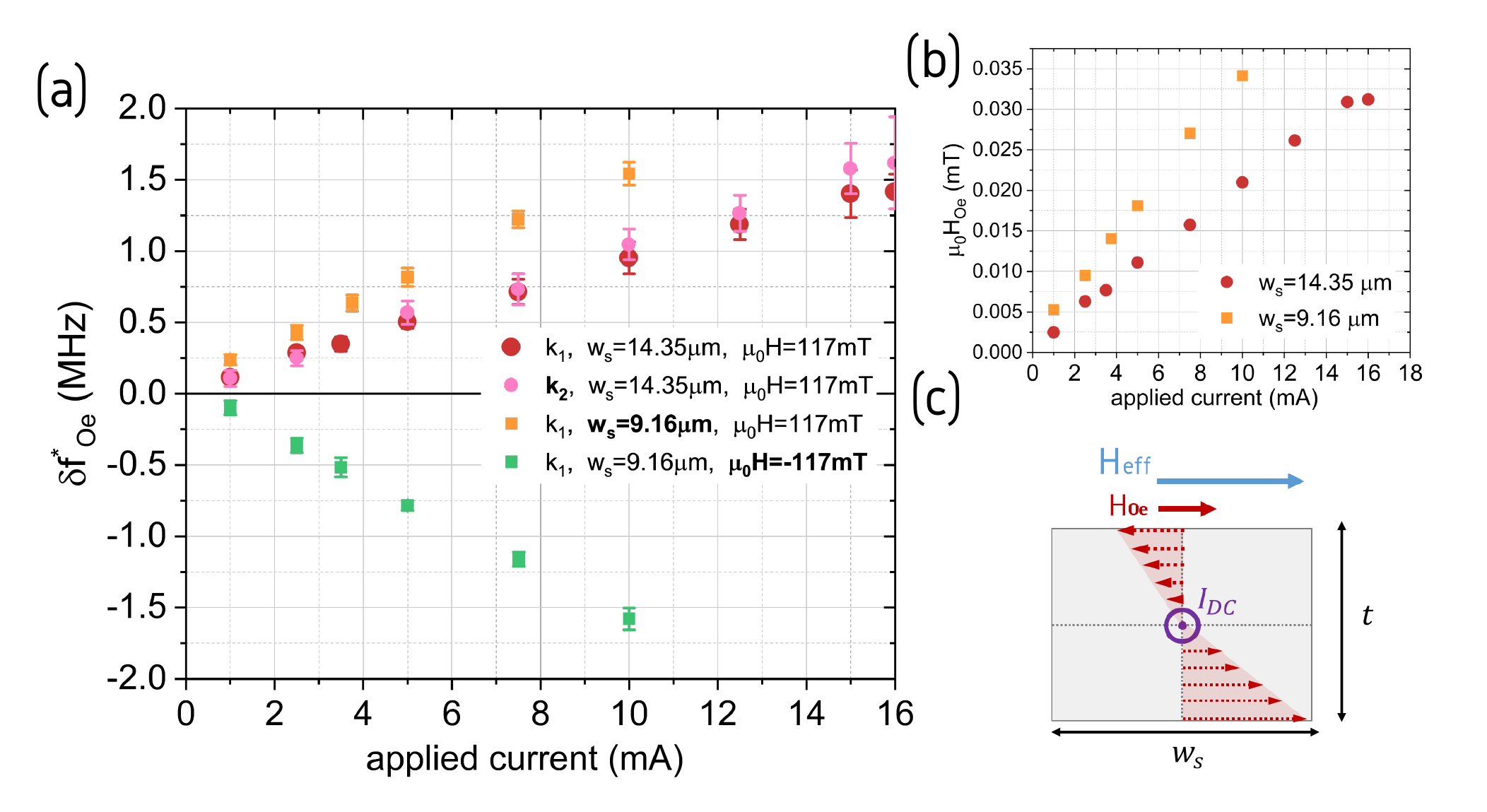}
\caption{Frequency shift due to the Oersted field in a patterned strip on a MgO/Co$_2$MnSi(20nm)/MgO/Ti film under an applied magnetic field $\mu_0 |H|=117~\text{mT}$. Here $k_1=3.9~\text{rad/}\mu\text{m}$ and $k_2=1.5~\text{rad/}\mu\text{m}$ are the two accessible wave vectors, $w_\text{s}$ is the nominal width of the strip. (a) Measurements of the Oersted frequency shift as a function of the applied current. (b) Estimation of the magnitude of the corresponding Oersted field obtained from the corresponding frequency shifts. (c) Sketch of the strip cross-section, the applied current and the geometry of the Oersted field corresponding to the measured shift. } 
\label{fig:Oersted}
\end{figure}

In Fig.~\ref{fig:Oersted} (a) we present the complementary current-induced frequency shift due to the Oersted field of the applied DC current $\delta f_\text{Oe}^*$. This also shows a linear behavior as function of the applied current. Additionally, it switches magnitude with applied magnetic field, is independent of the probed wave vector $k$, and reduces its magnitude for wider strips. In Fig.~\ref{fig:Oersted} (b) we present an estimation of the Oersted field responsible for this shift. We calculate this field using the resonance frequency $\delta f_\text{Oe}= \frac{\partial f}{ \partial H} H_\text{Oe}.$ In Fig.~\ref{fig:Oersted} (c) we have sketched a qualitative distribution of the Oersted field in a cross section of the material that would lead to the calculated sign of the field. Note that if we normalize the magnitude of $\delta f_\text{Oe}$ by the width of the Co$_2$MnSi strip $w_\text{s}$, we obtain the same magnitude showing that this shift depends only on the current density as expected.

\subsection{Non-adiabatic STT parameter analysis}
The non-adiabatic STT produces current-induced dissipative losses \cite{Bailleul.2017} and in turn will modify the spin wave amplitude. In our measurements, this amplitude is proportional to the mutual inductance between the antennas $\Delta L_{ij}$. Therefore, any spin wave amplitude change produced by the application of an electric current and the non-adiabatic STT should show up in this experimental observable too. Although this is an intuitive conclusion, relating the actual damping coefficients with the measured spin wave amplitude requires considering the magnetization precession and its dissipation. 

It can be shown that in the presence of magnetic damping the susceptibility of a ferromagnet will have a finite field linewidth at resonance $\delta H = 2\frac{\alpha \omega_\text{res}}{\gamma \mu_0}$ \cite{Vlaminck.2008}. This can also be seen as a frequency linewidth since $\delta \omega = \frac{\partial \omega}{\partial H} \delta H$. This means that the magnetic losses can be associated with a frequency linewidth $\delta \omega$, i.e. there is a finite time ($\propto 1/\delta \omega$) in which the loss of energy occurs. $T_2$ is defined as a characteristic time in which the relaxation due to the damping takes place:

\begin{equation}
\begin{aligned}
    \label{eq:T2}
    \frac{2\pi}{T_2} = \delta \omega =\frac{ \alpha \omega }{ \mu_0 \gamma } \frac{\partial \omega}{\partial H}.
\end{aligned}
\end{equation}

This characteristic time describes the exponential decay of the amplitude of the spin waves and therefore also of $\Delta L_{ij}$:

\begin{equation}
\begin{aligned}
    \label{eq:deltaLAndTime}
    |\Delta L_{ij}| (t_p=1/f_p) \propto e^{-1/(f_pT_2)},
\end{aligned}
\end{equation}

where $t_p=1/f_p$ is the apparent delay time it takes for spin waves to go from the excitation antenna to the measurement antenna.

As we have explained, non-adiabatic STT (parameterized by $\beta$) formally plays the same role as the intrinsic magnetic damping (parameterized by $\alpha$) and can be considered as an additional damping term [see torque equation, Eq.~(1), in the main text]. Therefore, under the influence of a spin-polarized current, the characteristic time in Eq.~\eqref{eq:T2} becomes \cite{Haidar.2012}:

\begin{equation}
\begin{aligned}
    \label{eq:T2STT}
    \frac{2\pi}{T_2} = \frac{ \alpha(\omega_\text{res}+\delta \omega_{ij}) - \beta \delta \omega_\text{Dop} }{ \mu_0 \gamma } \frac{\partial \omega}{\partial H}.
\end{aligned}
\end{equation}

Note that we have included the current-induced frequency shift $\omega_\text{k} \rightarrow \omega_\text{k}+\delta \omega_{ij}$ [see Eq.~\eqref{eq:shiftij}].

Now, replacing Eq.~\eqref{eq:T2STT} into Eq.~\eqref{eq:deltaLAndTime}, give us the attenuation coefficient for a spin wave travel during time $1/f_\text{p}$ (along the distance $D$) between the antennas under the influence of a spin polarized current. If we compare this to our parameterization of the current-modified mutual inductance [Eq.~\eqref{eq:mutualInductancesSTT}], we can identify:

\begin{equation}
\begin{aligned}
    \label{eq:epsilon&Beta1}
    |\Delta L_{ij}(f,\pm I)| \propto \left( 1 \pm \varepsilon_{ij} (\pm I) \right) A \propto e^{-\frac{\alpha (\omega_\text{k} \pm \delta \omega_{ij}) -\beta \delta \omega_\text{Dop}}{2\pi \gamma \mu_0 } \frac{\partial \omega}{\partial H} \frac{1}{f_\text{p}}}.
\end{aligned}
\end{equation}
 
Since $\varepsilon_{ij}$ parameterizes the current-induced change in amplitude, we have the following.

\begin{equation}
\begin{aligned}
    \label{eq:epsilon&Beta2}
    1 \pm \varepsilon_{ij} (\pm I) & = e^{ -\frac{ \pm \alpha \delta \omega_{ij} -\beta \delta \omega_\text{Dop}}{2\pi \gamma \mu_0 } \frac{\partial \omega}{\partial H} \frac{1}{f_\text{p}} },\\
   \pm \varepsilon_{ij} (\pm I) \approx ln\left(1 \pm \varepsilon_{ij} (\pm I) \right)& = -\frac{1}{2\pi \gamma \mu_0 f_\text{p}} \frac{\partial \omega}{\partial H} \left( \pm 
\alpha \delta \omega_{ij}- \beta  \delta \omega_\text{Dop} \right),
\end{aligned}
\end{equation}

assuming that $\varepsilon_{ij} \ll 1$. 

Now, considering the two propagation directions and their respective frequency shifts [Eqs.\eqref{eq:shift21} and \eqref{eq:shift12}]:

\begin{subequations}
\label{eq:epsilons}
\begin{eqnarray}
\varepsilon_{21}(+I)=\varepsilon_{21} \approx  -\frac{1}{ \gamma \mu_0 f_{\text{p},21}} \frac{\partial \omega}{\partial H} \left[ \alpha \delta f_\text{Oe}^* + \alpha \delta f_\text{NROe}^* + (\alpha-\beta)\delta f_\text{Dopp}(I)^*  \right],\label{subeq:epsilon21}\\
\varepsilon_{12}(+I)=\varepsilon_{12} \approx  -\frac{1}{ \gamma \mu_0 f_{\text{p},12}} \frac{\partial \omega}{\partial H} \left[ \alpha \delta f_\text{Oe}^* - \alpha \delta f_\text{NROe}^*- (\alpha-\beta)\delta f_\text{Dopp}(I)^*  \right].\label{subeq:epsilon12}
\end{eqnarray}
\end{subequations}

If we define:

\begin{equation}
\begin{aligned}
    \label{eq:Factor}
    T_{ij}= -\frac{2}{ \gamma \mu_0 f_{\text{p},ij}} \frac{\partial \omega}{\partial H},
\end{aligned}
\end{equation}

we can write:

\begin{subequations}
\label{eq:SumDiffEpsilons}
\begin{align}
Q_\text{anti} = \frac{\varepsilon_{21}}{T_{21}} - \frac{\varepsilon_{12}}{T_{12}}  =& \: (\alpha-\beta)\delta f_\text{Dopp}(I)^*,\label{subeq:alphaBeta}\\
Q_\text{sym} = \frac{\varepsilon_{21}}{T_{21}} + \frac{\varepsilon_{12}}{T_{12}} =& \: \alpha \delta f_\text{Oe}(I)^*.\label{subeq:alphadOe}
\end{align}
\end{subequations}

where we have neglected the term $\alpha \delta f_\text{NROe}^*$ in Eq.~\eqref{subeq:alphaBeta} since $\alpha \ll 1$ and $\delta f_\text{NROe}^*$ is negligible as explained in part~\ref{subsecA}. From our previous analysis, we already know $\delta f_\text{Dopp}^*$, $\delta f_\text{Oe}^*$, $f_{\text{p},ij}$, therefore, we can obtain $\alpha-\beta$ using Eqs.~\eqref{eq:epsilons} or~\eqref{subeq:alphaBeta}. Note that the fact that $\alpha$ and therefore $\alpha\delta f_\text{Oe}^*$ are small allows us to experimentally determine the component $(\alpha-\beta)\delta f_\text{Dopp}(I)^*$ to $\varepsilon_{ij}$ reliably [see Eqs.~\eqref{eq:epsilons}].

\begin{figure}[H]
\centering
\includegraphics[width=0.65\textwidth]{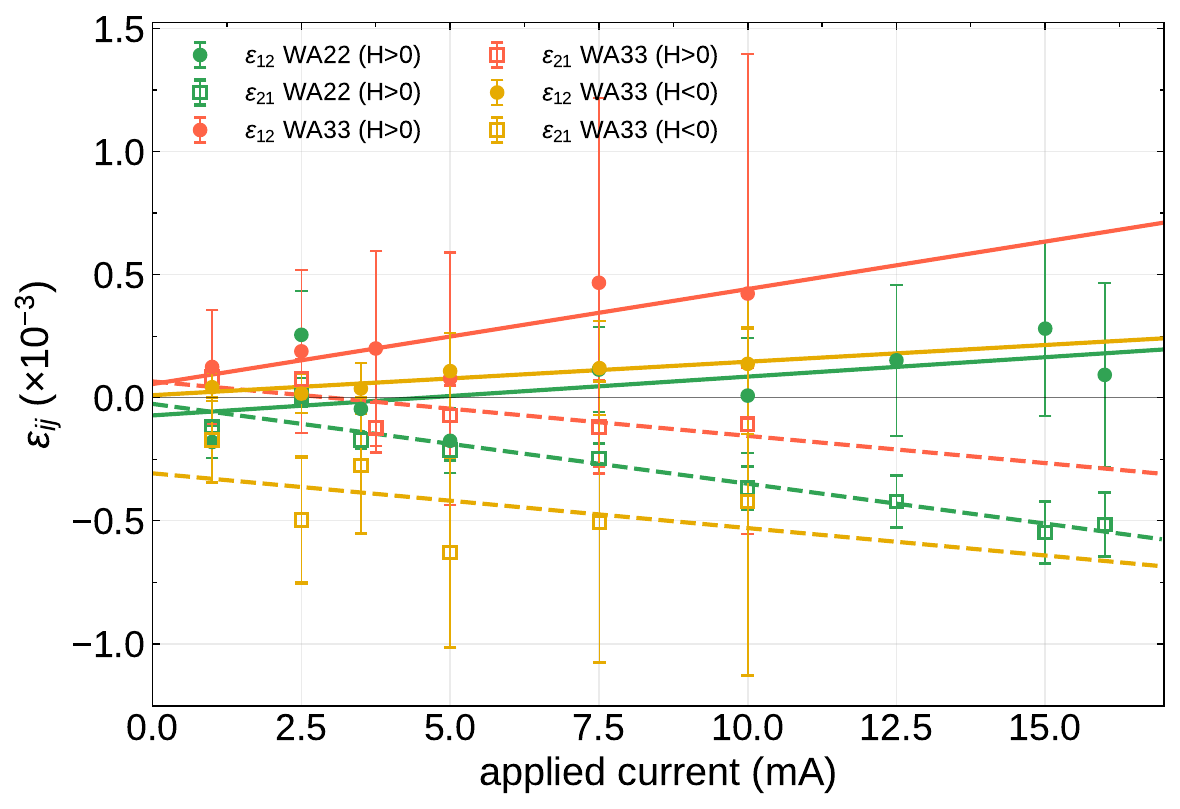}
\caption{Current-induced relative change of amplitude $\varepsilon_{ij}$ as function of the applied electric current for devices WA22 (presented in main text) and WA33 at an external magnetic field of magnitude 117 mT and wave vector $k_1=3.9~\text{rad/}\mu\text{m}$. For the case of WA33 we present the recorded data for the two field directions. The lines are the corresponding linear fits.} 
\label{fig:Epsilons}
\end{figure}

\begin{figure}[H]
\centering
\includegraphics[width=0.85\textwidth]{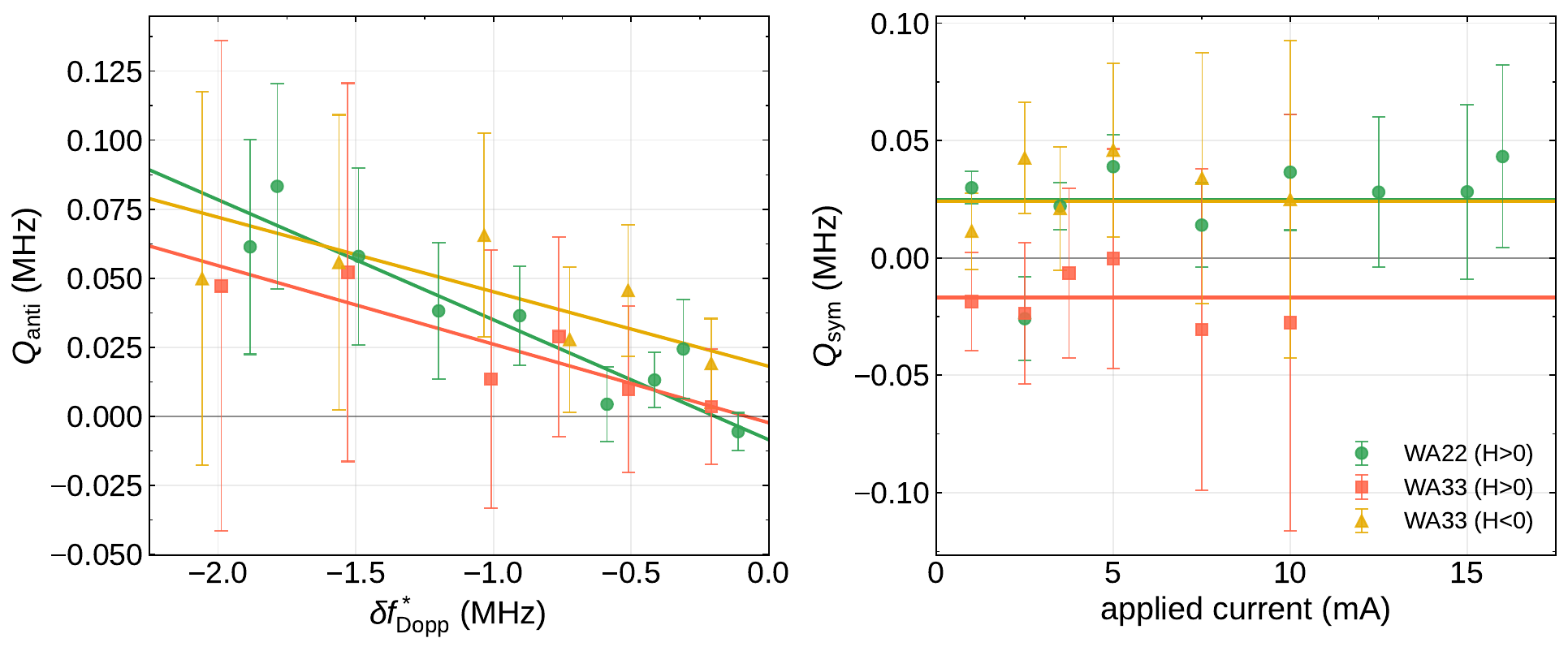}
\caption{ (left) $Q_\text{anti}$ as function of the spin-wave Doppler shift $\delta f_\text{Dopp}(I)^*$, the lines are fits to Eq.~\eqref{subeq:alphaBeta}. (right) $Q_\text{sym}$ as function of the applied electric current, the lines are just fits to constant values. The data corresponds to devices WA22 and WA33 at an applied magnetic field of magnitude 117 mT and wave vector $k_1=3.9~\text{rad/}\mu\text{m}$. For the case of WA33 we present the recorded data for the two field directions.} 
\label{fig:Qs}
\end{figure}

Experimentally we obtain the parameters $\varepsilon_{ij}$ by fitting the real part of $R_{ij}$ to Eq.~\eqref{eq:ReRij} [Fig.~\ref{fig:Rij} (b)]. We plot the resulting $\varepsilon_{ij}$ as functions of the applied electric current in Fig.~\ref{fig:Epsilons}. We can observe that $\varepsilon_{ij}$ varies linearly with the applied current. Also, we observe that there is a change of sign for the $\varepsilon_{ij}$ corresponding to the opposite direction of propagation of the spin waves. This sign reversal is what one expects from a non-adiabatic STT contribution to the spin wave attenuation [see Eqs.~\eqref{eq:epsilon&Beta2}]. In addition, there are vertical and horizontal offsets for some of the $\varepsilon_{ij}$. These are artificially introduced by our analysis and extraction due to the uncertainty that the oscillations in $R_{ij}$ introduce (see Fig.~\ref{fig:Rij}). This is also reflected in the large error bars we present in Fig.~\ref{fig:Epsilons}. Nevertheless, we focus on the change of $\varepsilon_{ij}$ with the applied electric current (i.e. slopes). We then proceed to calculate the quantities $Q_\text{anti}$ and $Q_\text{sym}$ [Eqs.~\ref{eq:SumDiffEpsilons}], plot them in Fig.~\ref{fig:Qs} and the results of their linear fits are presented in Table~\ref{tab:Q_fits}.

We observe that $Q_\text{anti}$ does indeed have a linear variation with the spin-wave Doppler shift across the different data sets, as expected from Eq.~\eqref{subeq:alphaBeta}. Furthermore, the linear fits have very small intercepts, suggesting that there are no important additional effects to the non-adiabatic STT contribution. Similarly, for $Q_\text{sym}$ we observe no real discernible slope in agreement with Eq.~\eqref{subeq:alphadOe} and the fact that $\alpha \delta f_\text{Oe}^*(I) < 2\times10^{-3} \text{MHz}$ is negligible for all explored currents and devices (see Fig.~\ref{fig:Oersted}). This is an additional check for no spurious current-induced contributions. The non-zero constant values from the fits of this symmetric component reflect the artificially introduced offsets we mentioned above for the $\epsilon_{ij}$. 

Finally, by computing the weighted (by error bar)average, we obtain $\alpha-\beta = -0.04\pm0.01$. Finally, the non-adiabatic STT parameter $\beta=0.04\pm0.01$ can be obtained by using our ferromagnetic resonance estimate of $\alpha = 1.2\pm0.1\times10^{-3}$.

\begin{table}[ht]
\centering
\caption{Fit results for $Q_\mathrm{anti}$ and $Q_\mathrm{sym}$.
$Q_\mathrm{anti}$ is fitted with a free intercept:
$Q_\mathrm{anti} = (\alpha-\beta)\,\delta f_\mathrm{Dopp}^* + b_\mathrm{anti}$.
$Q_\mathrm{sym}$ is fitted as a constant offset $b_\mathrm{sym}$ (see Fig.~\ref{fig:Qs}).}
\label{tab:Q_fits}
\begin{tabular}{| c | c | c | c |}
    \hline
  & $(\alpha-\beta)$
  & $b_\mathrm{anti}$ (MHz)
  & $b_\mathrm{sym}$ (MHz) \\
\hline
WA22 ($H>0$) & $-0.043 \pm 0.011$ & $-0.009 \pm 0.007$ & $+0.025 \pm 0.005$ \\
WA33 ($H>0$) & $-0.028 \pm 0.034$ & $-0.002 \pm 0.023$ & $-0.017 \pm 0.014$ \\
WA33 ($H<0$) & $-0.027 \pm 0.026$ & $+0.018 \pm 0.018$ & $+0.024 \pm 0.011$ \\
\hline
\end{tabular}
\end{table}

\bibliography{HeuslerP_sup}
\bibliographystyle{apsrev4-2}